\documentclass[aps,prl,twocolumn,groupedaddress,dvipdfmx]{revtex4-1}
\usepackage{amsmath}
\usepackage{amssymb}
\usepackage{bm}
\usepackage{dcolumn}
\usepackage{graphicx}
\usepackage{mciteplus}
\usepackage{tabularx}
\usepackage{color}
\usepackage{url}
\usepackage{threeparttable}

\begin{document}
\title{A computational scheme to evaluate Hamaker constants of molecules with practical size and anisotropy}
\author{Kenta Hongo}
\email{kenta\_hongo@mac.com}
\affiliation{School of Information Science,
  JAIST, Asahidai 1-1, Nomi, Ishikawa 923-1292, Japan}
\altaffiliation{National Institute for Materials Science (NIMS), 
1-2-1 Sengen, Tsukuba, Ibaraki 305-0047, Japan}
\altaffiliation{PRESTO, JST, 
4-1-8 Honcho, Kawaguchi, Saitama 332-0012, Japan}
\author{Ryo Maezono}
\affiliation{School of Information Science,
  JAIST, Asahidai 1-1, Nomi, Ishikawa 923-1292, Japan}

\date{\today}

\begin{abstract}
We propose a computational scheme to evaluate 
Hamaker constants, $A$, of molecules with 
practical sizes and anisotropies. 
Upon the increasing feasibility of diffusion 
Monte Carlo (DMC) methods to evaluate binding curves 
for such molecules to extract the constants, 
we discussed how to treat 
the averaging over anisotropy and 
how to correct the bias due to the non-additivity.
We have developed a computational procedure for dealing with the anisotropy
and reducing statistical errors and biases in DMC evaluations, based on
possible validations on predicted $A$.
We applied the scheme to 
cyclohexasilane molecule, Si$_6$H$_{12}$, 
used in 'printed electronics' fabrications, 
getting $A \sim 105 \pm 2$ [zJ], being 
in plausible range supported even by 
other possible extrapolations. 
The scheme provided here would open a 
way to use handy {\it ab initio} evaluations 
to predict wettabilities as in the form 
of materials informatics over broader molecules. 
\end{abstract}
\maketitle

\section{Introduction}
Hamaker constants{~\cite{1937HAM}}, $A$, 
dominate the wettability{~\cite{2009BON,2011ISR}}
of solvents, which is one of the critical properties in
industrial applications of Sol-Gel methods{~\cite{2015LEV}},
including solution processes for semiconductor devices.{~\cite{2006SHI}} 
Microscopic insights on the wettability{~\cite{2009BON,2011ISR}} 
relates the Hamaker constant with molecular interactions, 
which can be, in principle, evaluated from {\it ab initio} 
simulations. 
From the asymptotic behavior of molecular binding curves, 
or potential energy surfaces (PES), $\sim C_6/R^6$, 
the Hamaker constant can be 
computed as $A_\mathrm{add} = \pi C_6 \rho^2$, 
provided that only a binding with a single 
$C_6$ matters and a naive superposition is 
expected.~\cite{2011ISR} 
The index, 'add', then stands for 'additive' 
and $\rho$ denotes the molecular density which 
appears when the superposition integral is counted. 
Though we can find several such prototypical 
works{~\cite{2011ISR}} 
of the '{\it ab initio} assessment' applied to simple and
highly symmetric molecules,
we would immediately encounter troubles when attempting
to apply the framework to practical solute molecules. 
Most molecules of industrial interest are not 
so highly symmetric that 
we cannot generally expect the additivity 
of the interaction.~\cite{2014MIS}
In these cases, too many alignments of coalescence
are possible due to the anisotropy of molecules,
bewildering us how to model the coalescence 
with the confidence for capturing the nature 
of the system. 

\par
The main subject of the present paper is 
how to estimate $A$ 
for the practical solute molecules 
via $A_\mathrm{add}$ with plausible 
considerations mainly for the anisotropy. 
Once we could establish such a scheme, 
such database of molecular 
interactions aided by recent {\it ab initio} 
methods~\cite{1997STO,2006KAP} 
can provide the Hamaker constants 
over various liquids. 
It would help to predict, control, and design
such solution processes including not only wettablities but also
suspensions and solvabilities by using empirical molecular dynamics
simulations.{~\cite{2009BON}} 

\par
The present study has been originally motivated by 
the demand to estimate $A$ for 
a cyclohexasilane molecule, Si$_6$H$_{12}$ (CHS), 
which is used as an ink for 'printed electronics' 
technology to fabricate 
polycrystalline Si film transistors.{~\cite{2006SHI}}
The ink including Si-based precursors is sprayed on a substrate,
which is sintered to form an amorphous Si thin film,
without using expensive vacuum equipment in the conventional
semiconductor processes.
The ink printing process has hence attracted recent interests for
realizing more saving and lower environmental impact 
technology.~{\cite{2006SHI}}
Controlling the wettability of these inks is of rather general interest
because the technology is about to be applied further to 
fabricate oxide or carbon nanotube film semiconductor
devices~\cite{2013LAU,2014INO} 
by using various inks instead of Si-based ones.
For going beyond conventional/experimental preparations of inks, 
several simulations have been made to analyze the wettability of droplets
on ink-jet processes dynamically using molecular
dynamics~\cite{2013NAK} or empirical models~\cite{2012MAT}.
The predictability of these simulations strongly depends
on the force fields that are currently prepared by empirical
parameterizations of Lennard-Jones type potentials.
The {\it ab initio} assessment for these parameterizations 
is obviously recognized as an important breakthrough 
in getting more universal applicability. 

\par
For CHS, there is no reference to $A$, and then 
we tried evaluating $A_\mathrm{add}$ from its binding curve. 
Besides the anisotropy discussed above, 
the commonly available framework, DFT (density functional theory), 
is known to fail to describe molecular interactions 
mostly, and the DFT performance strongly depends on 
exchange-correlation (XC) functionals adopted.{~\cite{2012COH}}
In the present case, the interaction of this system,
CHS, is of non-$\pi$ staking nature, known as
an {\it aliphatic-aliphatic} one{~\cite{2011KIM}}
between the $\sigma$ bonds at the 
HOMO (highest-occupied molecular orbital) 
levels of the monomers.
Unlike {\it aromatic-aromatic} interactions of {\it e.g.} benzene dimer,
there has been only a few investigations
on {\it aliphatic-aliphatic} interactions
and hence no established scheme of how to treat
the anisotropy of molecules in the evaluation of binding curves
even for moderately tractable size and symmetry of the target molecules.
As is well-known, accurate correlated methods such as CCSD(T)
are required to get enough reliable estimations of molecular
interaction.{~\cite{2009HOB,2012COH}}
Such methods are, in general, quite costly in the sense of
the scalability on the system size $N$, {\it e.g.},
$\sim N^7$ for CCSD(T){~\cite{2000HEL}}.
Such severe scalabilities obstruct the applications to
larger molecules being likely in the practical cases.
In contrast, DMC (diffusion Monte Carlo) method is quite promising
and its applicability to more practical issues gets rapidly
extended.~\cite{2010NEE,2012AUS,2011UEJ,2013UEJ} 
This framework is regarded in principle as the most reliable
that can achieve 'numerically exact solutions' in some cases~\cite{2006HON,2007HON},
and there has been so far several applications to noncovalent 
systems{~\cite{2008KOR,2013HOR,2014HOR,2013DUB,2014DUB,2013HON,2010HON,2012WAT,2015HON,2016HON,2016DUB}},
to calibrate even over accurate molecular orbital methods such as CCSD(T).
DMC scales at worst to $\sim N^3$,{~\cite{2012AUS}} making it possible to
be applied further to larger molecules including molecular 
crystals.{~\cite{2013HON,2010HON,2012WAT,2015HON,2016HON}}

\par
In this paper, we therefore applied DMC to evaluate 
$A_\mathrm{add}$ of CHS. 
Upon a careful benchmark on benzene molecule (given in Appendix C), 
we have established a scheme 
(i) coping with the anisotropy of the molecules, 
(ii) reducing statistical errorbars and biases
that are small enough for a usable predictions, and
(iii) based on several possible validations on 
the predicted $A$ for which no experimental 
reference value is available. 
The scheme is applied to CHS getting  
$A = 105 \pm 2$ [zJ] which is in a reasonable range 
validated by several side considerations. 
By making comparisons with binding curves by DFT, 
we also provide a useful calibration over 
several XC for the predictability of $A$. 

\par
The paper is then organized as follows: 
In the main body of the paper, we provide 
descriptions of the scheme applied to CHS, 
followed by validations of the prediction 
as briefly as possible so as to concentrate 
on following the established procedure. 
Thus, put aside into appendices are 
detailed descriptions for computational methods
(Appendix A), some formalism of Hamaker constants
considered in the present work (Appendix B),
and all the discussions on the validations of the 
procedure made on the benzene dimer benchmark (Appendix C-E).
Technical details about evaluation of $A_\mathrm{add}$
for CHS are also given alongside the benzene case in the appendices.
Summaries of the paper are given as Concluding Remarks
at the end of main text.
For detailed correction schemes, such as BSSE
(basis set superposition error),
CBS (complete basis set) schemes as well as time-step 
error in DMC are given in Supporting Information.

\section{Results and Discussions}
\subsection{Hamaker constant of CHS}
\label{results}
To get $C_6$, we evaluated dimer binding curves of CHS 
for three types of coalescences, {\it i.e.}, 
Sandwich (Type-A), T-shape (Type-B), 
and Parallel (Type-C), as shown in 
Fig.{~\ref{dimer-pattern}. 
Computational details for the evaluation are
given in Appendix A. 
For the CHS monomer structure, 
we took the chair conformation 
{~\cite{1994LEO,2006KOR}} since it is 
known to be most stable. 
The monomer geometry is optimized at
the B3LYP/6-311G level using Gaussian09.{~\cite{2009FRI}}
To plot a binding curve,
we vary binding distances of a dimer coalescence, keeping
each of the monomer structures fixed to the above one.
This is valid to some extent because
we focus on $C_6$ extracted from the long-range behavior
where each of the monomer structures may be almost 
the same as that of an isolated monomer.
The inter-monomer distance is defined as that between 
the centers of gravity of the monomers. 

\begin{figure}[htb]
\begin{center}
  \includegraphics[scale=0.6]{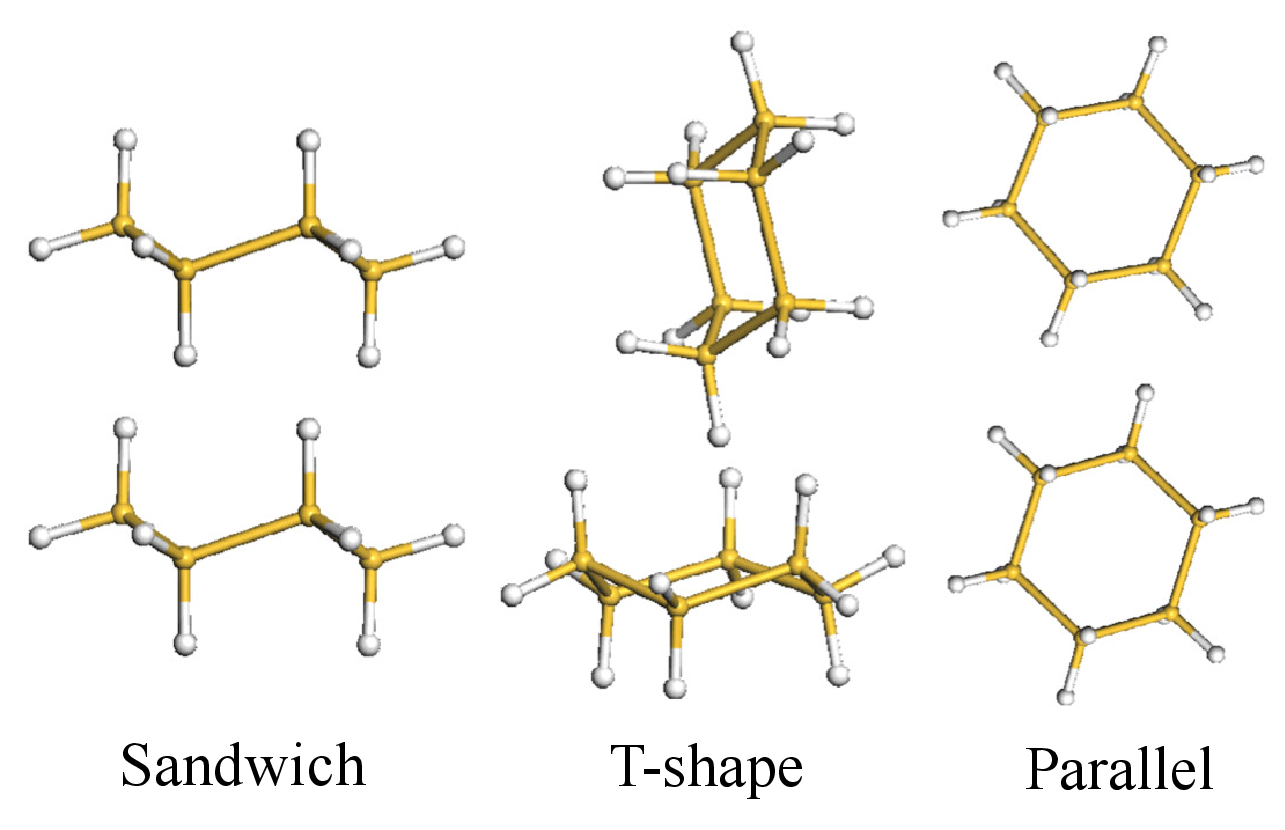}
\end{center}
\caption{
  Typical configurations of the dimer coalescence
  considered in this work, 
Sandwich (Type-A), T-shaped (Type-B), 
and Parallel (Type-C).}
\label{dimer-pattern}
\end{figure}

\par
Fig.~\ref{ljfitting} shows DMC binding curves for each coalescence 
configurations, compared with CCSD(T) references.
The sandwich (Type A) configuration is identified to give 
the most stable binding energy $\Delta E$ over the others, 
$p\sim \exp(-\Delta E/kT) \sim $ 98 \% at $T = 298.15$K
as given in Table~\ref{table:two-body}. 
Our careful benchmark for benzene case given in Appendix C clarifies 
that the deepest binding configuration almost dominates 
Hamaker constants. 
We can therefore concentrate only on 
the Type A binding curve to extract $C_6$ from its asymptotic behavior.
Hereafter we adopt a symbol, $C_6^{\rm stable}$,
as a $C_6$ value for the most stable coalescence configuration.
\begin{table*}[htb]
\begin{tabular}{ccccc}
  &  B3LYP-GD3 & MP2 & CCSD(T) & DMC \\ [2pt]
\hline
$p(\mathrm{A})$/$R_{\rm e}$  & $0.971/4.9$ & $0.976/4.7$
& $0.968/4.8$ & $0.987(82)/4.83(2)$ \\ [2pt]
$p(\mathrm{B})$/$R_{\rm e}$  & $0.026/6.6$ & $0.023/6.1$
& $0.030/6.2$ & $0.010(24)/6.37(6)$ \\ [2pt]
$p(\mathrm{C})$/$R_{\rm e}$  & $0.003/8.8$ & $0.002/8.4$
& $0.003/8.6$ & $0.003(14)/8.7(1)$ \\ [2pt]
$\bar{R}_{\rm dim}$            & $5.0$       & $4.7$       & $4.9$
& $4.9(1)$ \\ [2pt]
\hline
\end{tabular}
\caption{
Comparisons of the equilibrium stability among 
three coalescence configurations in 
Fig.~\ref{dimer-pattern}, in terms of the 
thermal probability weight, $p \sim \exp (-\Delta E/kT)$.
Equilibrium binding lengths, $R_\mathrm{eq}$ [\AA{}]),
are also shown.
$\bar R_{dim}$ [\AA{}] is the thermal averaged  binding lengths 
at $T = 298.15$ K for each method.
}
\label{table:two-body}
\end{table*}

\begin{figure}[htb]
\begin{center}
  \includegraphics[scale=0.80]{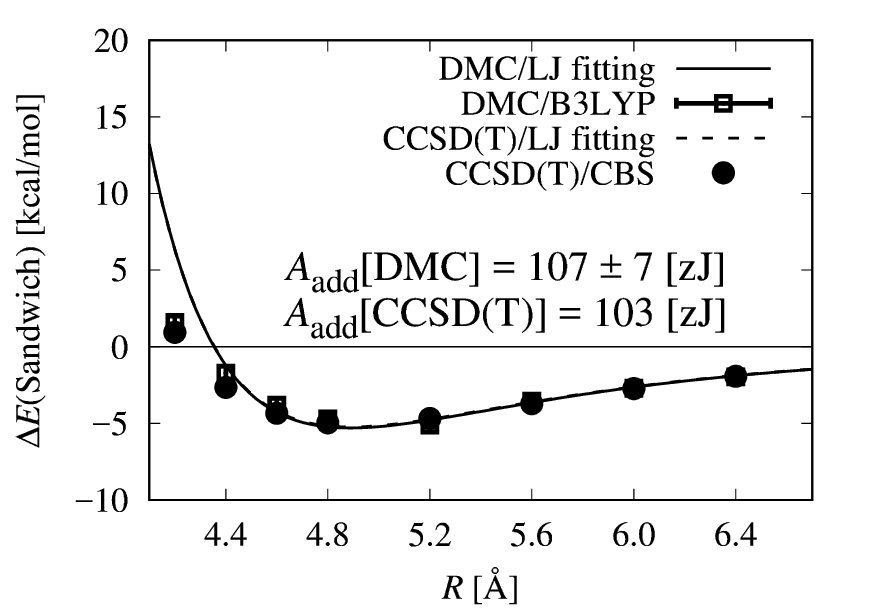}
  \includegraphics[scale=0.80]{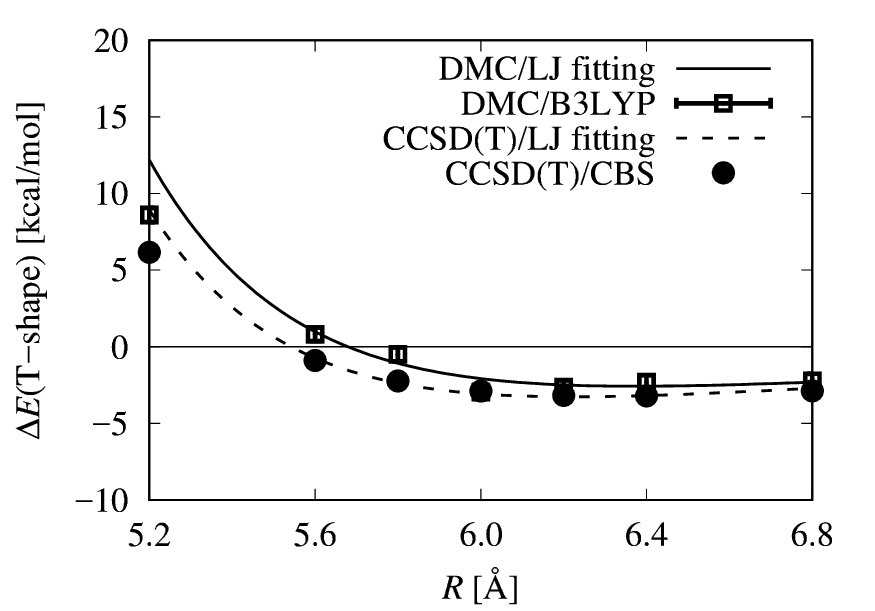}
  \includegraphics[scale=0.80]{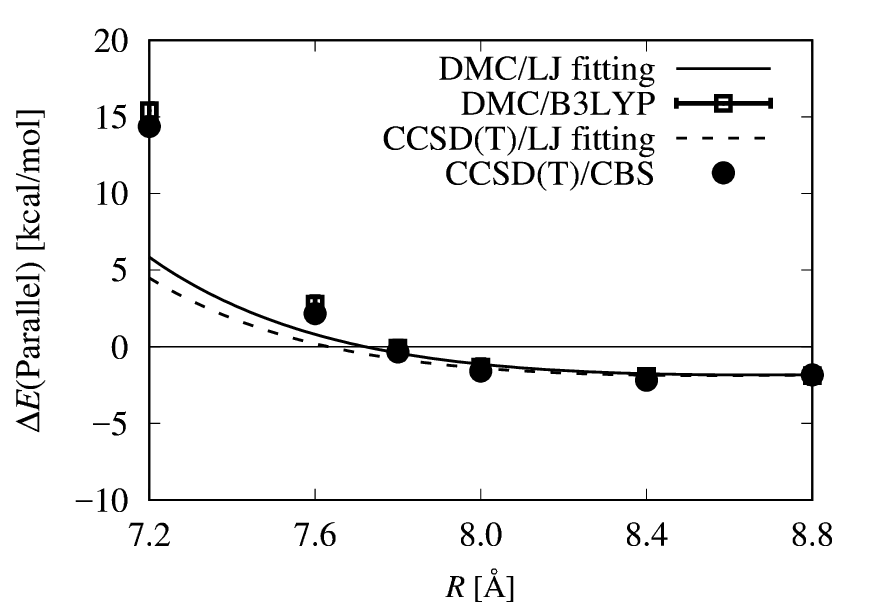}
\end{center}
~\caption{
  DMC binding curves for three types of coalescence (Sandwich/T-shape/Parallel) 
  compared with CCSD(T). For eye-guide, Lennard-Jones fitting is depicted
  for both DMC and CCSD(T), though for T-shape and Parallel, 
  the fitting has a limited meaning because they do not behave like $R^{-6}$
  (see Appendix D for details)}.
\label{ljfitting}
\end{figure}
\par
To extract $C_6^{\rm stable}$ from PES data, 
we considered several fitting schemes: 
log-fit, 6-12 Lennard-Jones (LJ), 
and the power-fit for the 
correlation energy defined as the deviation from
Hartree-Fock energy (denoted as $C_6^{\rm stable;LOG},
C_6^{\rm stable;LJ}$, and $C_6^{\rm stable;Corr}$).
The estimation is also affected by the 
choice of which distance range is taken to be fit. 
We make detailed discussions of our fitting schemes in
Appendix D, considering the benzene benchmark
as well as CHS.
From them, we find out that the power-fit for the 
correlation energy~\cite{1937LON,2016GRI} 
\begin{equation}
  \label{disp}
\Delta E_\mathrm{corr}(R) = -\frac{C_6}{R^6} -\frac{C_8}{R^8} \cdots ,
 \end{equation}
achieves small enough errorbars for usable predictions on 
the Hamaker constant by DMC, 
$A_\mathrm{add}(C_6^{\rm stable;Corr}) = 105 \pm 2$ [zJ], 
with the fitting range $R = 4.4\sim 6.4$\AA{} which 
minimizes a measure of the deviation from the fitting model (Appendix D), 
where the experimental density of CHS, $\rho = 0.323 \times 10^{28}$[1/m$^3$] 
at 298.15K.~\cite{2001CHO} was used to evaluate $A_\mathrm{add}=\pi\rho^2 C_6$.
We note that this scheme is applicable 
only to many-electron wavefunction methods
such as DMC, CCSD(T), and MP2.
The Hamaker constants evaluated from various approaches 
(methods/schemes) in the present study
are listed in Table~\ref{table:physicalquantities},
and their validation is given in the next subsection.

\subsection{Validation of $A$ value}
We found that our DMC evaluations of $A_\mathrm{add}$
agree with those obtained from
the other reliable quantum chemistry methods,
CCSD(T) and MP2,
implying our {\it ab-initio} evaluation schemes
would be reasonable within the framework of
many-electron wavefunction theory.
But, there is no reference to $A$ to be compared
directly to the present estimation for CHS.
So we tried a validation via side-way manner as follows:
(1) A simple estimation using London's theory~\cite{1937LON} would give 
underestimated reference as discussed in Appendix C.
Static polarizabilities and ionization energies
can be evaluated using HF and B3LYP levels of theory to give $C_6$ values 
(denoted as $\langle C_6^\mathrm{London} \rangle^\mathrm{iso}$) 
and then $A_\mathrm{add}(\langle C_6^\mathrm{London} \rangle^\mathrm{iso}) = 66$ 
and $81$ [zJ], respectively 
($\langle \dots \rangle^\mathrm{iso}$ means an isotropic orientation average,
described in Appendix B and C).
The values are consistent in the sense that 
they are actually located in the underestimated 
range compared with the other estimations in 
Table~\ref{table:physicalquantities}.
\par
(2) As another trial for the validation
using the estimations ($A_\mathrm{L}$) based on 
the Lifshitz theory~\cite{1956LIF} 
whose formalism is given in Appendix B, 
we consider the dependence on the molecular 
weights of $A\propto C_6$. 
Since the dispersion interactions scale 
to the total polarization, 
it is not so bad expectation that $A$ 
is roughly proportional to molecular weights. 
Under this assumption, the ratio, 
$A_\mathrm{L}$(C$_6$H$_{12}$)/$A_\mathrm{L}$(C$_5$H$_{10}$) 
= (53.0$\pm$0.2)/(49.4$\pm$0.3), 
can be taken as being equal to 
$A_\mathrm{L}$(Si$_6$H$_{12}$)/$A_\mathrm{L}$(Si$_5$H$_{10}$). 
Using the known value of $A_\mathrm{L}$ (73.4 $\pm$ 0.4 zJ) 
for CPS(Si$_5$H$_{10}$), we can roughly estimate 
that of CHS as $A_\mathrm{L}$(extrapol.)=78.9 $\pm$ 0.5 [zJ]. 
Another possible regression can be made in terms of 
$C_6$ instead of $A$. 
Regressing the quadratic functions to
the TDDFT and EMT (effective medium theory)
data on $C_6$ values of the Si$_n$H$_m$ family~\cite{2008BOT}
(denoted as $C_6^{\rm extrapol.}$),
we get $A_\mathrm{add} (C_6^{\rm extrapol.}) = 110$ [zJ] and 
$94$ [zJ], respectively. 
These values lie within a reasonable range by comparison 
with those in Table~\ref{table:physicalquantities}, 
being consistent with the fact that 
$A_\mathrm{add}$ is larger than $A_\mathrm{L}$(extrapol.).
We found the EMT extrapolation of $A_\mathrm{add}$ 
being closer to $A_\mathrm{L}$(extrapol.).
This may be attributed to the fact that, in EMT, 
$C_6$ is evaluated by the dielectric constants 
modelled by those of bulk quantities, 
as in the Lifschitz theory. 
The reason why $A_\mathrm{add}$ generally overestimates
$A$ values compared with $A_\mathrm{L}$ may be explained as follows:
$A_\mathrm{add}(C_6^{\rm stable})$ is evaluated only from the 
longest-ranged exponent with a selected coalescence 
configuration, Type-A in the present case.
The other configurations with shorter-ranged exponents
should be included in liquids by some fractions,
and hence effectively weaken the binding strength 
estimated under such an assumption with 100\%
constitution of Type-A coalescence. 
Such an effect would be 
represented as 'effectively reduced' Hamaker 
constants close to $A_\mathrm{L}$.
Hence, the values, $A_\mathrm{add}/A_\mathrm{L}$,
could be sorted out by 
a factor dominating the fraction, 
$\exp{\left(-\Delta E/kT\right)}$, 
where $\Delta E$ denotes a typical energy difference 
between the coalescence configurations with 
the longest- and the shortest-ranging exponents.

\begin{table*}[htb]
  \begin{tabular}{ccccc}
    Method/Scheme & 
    LOG & 
    LJ  &
    Corr &
    London \\ [2pt]
    \hline
LDA       & $48$       & $90$     &          &       \\ [2pt]
M06-2X    & $36$       & $56$     &          &       \\ [2pt]
B3LYP     &            &          &          & $81$\footnotemark[1] \\ [2pt]
B3LYP-GD2 & $57$       & $62$     &          &       \\ [2pt]
B3LYP-GD3 & $96$       & $105$    &          &       \\ [2pt]
B97-D     & $98$       & $81$     &          &       \\ [2pt]
HF        &            &          &          & $66$\footnotemark[2]  \\ [2pt]
MP2       & $104$      & $99$     & $104$    &       \\ [2pt] 
CCSD(T)   & $95$       & $103$    & $106$    &       \\ [2pt]
DMC       & $99(30)$   & $107(7)$ & $105(2)$ &       \\ [2pt] 
\hline
\end{tabular}
\caption{
  Computed Hamaker constants $A_\mathrm{add}(C_6^\mathrm{stable})$ [zJ],
  based on different $C_6$ evaluation schemes,
  $C_6^\mathrm{stable;LOG}/C_6^\mathrm{stable;LJ}/
  C_6^\mathrm{stable;Corr}$
(see text for more details about the definitions).
Since HF and B3LYP give repulsive PESs,
their Hamaker constants cannot be evaluated by the present
PES scheme,
and the London scheme is used to estimate
$A_\mathrm{add}(\langle C_6^\mathrm{London}\rangle^\mathrm{iso})$,
instead.
Statistical errors in the DMC values are given in parenthesis. 
}
\label{table:physicalquantities}
\end{table*}
\footnotetext[1]{Ionization energy of $9.2$ eV and static
  polarizability of $179$ bohr$^3$ obtained at the B3LYP/BDF
  level were used to evaluate eq.~\eqref{london} in Appendix B.}
\footnotetext[2]{Ionization energy of $8.8$ eV and static
  polarizability of $165$ bohr$^3$ obtained at the HF/BDF
  level were used to evaluate eq.~\eqref{london} in Appendix B.}

\subsection{Validation of equilibrium properties}
\label{discussion}
Although the long-range behavior of PES
concerns with the evaluation of $A$,
validation of PES at equilibrium distance may also
give us some confidence in our numerical results.
Equilibrium properties including
binding energies ($\Delta E)$ and equilibrium lengths ($R_\mathrm{eq}$)
are summarized in Table~\ref{table:ng_table}.
The estimated binding energies in our DMC-PES
are comparable with the typical 
value of non-$\pi$ stacking energies
$\sim -5$ kcal/mol.~\cite{2011KIM}
Compared with $\pi$-stacking energies, it is about
twice larger, which would be consistent with the 
higher boiling temperature of CHS than that of
its structural isomers with the same molecular weights 
but without hydrogen bindings.~\cite{1997GRE}

\begin{table*}[htb]
\begin{tabular}{cllll}
  &  \multicolumn{2}{c}{Equilibrium properties} &  Short-range
  &  Long-range \\ \cline{2-3}
  & \multicolumn{1}{c}{$\Delta E(R_\mathrm{e})$}
  & \multicolumn{1}{c}{$R_\mathrm{e}$} & \multicolumn{1}{c}
  {$\Delta E(4.2)$}
  & \multicolumn{1}{c}{$\Delta E(7.0)$} \\
\hline
LDA        & $-9.39$[NG]  & $4.34$[NG]     &  $-8.73$[NG] &  $-0.44$[NG] \\ 
B3LYP-GD2  & $-3.47$[NG]  & $4.93$[G]      &  $2.09$[NG]  &  $-0.71$[NG] \\
B3LYP-GD3  & $-5.06$[G]   & $4.93$[G]      &  $5.74$[NG]  &  $-1.13$[G] \\
B97-D      & $-4.13$[NG]  & $4.88$[G]      &  $2.73$[NG]  &  $-1.24$[G] \\
M06-2X     & $-3.75$[NG]  & $4.67$[NG]     &  $1.84$[NG]  &  $-0.40$[NG] \\
MP2        & $-6.36$[NG]  & $4.70$[NG]     &  $-0.88$[NG] &  $-1.23$[G]  \\
DMC/B3LYP  & $-5.3(2)$[G] & $4.89(2)$[G]   &  $1.6(4)$(G] &  $-1.2(4)$[G] \\
CCSD(T)    & $-5.24$      & $4.89$         &  $0.94$      &  $-1.13$  \\
 \hline
\end{tabular}
\caption{
Summary of the abilities in describing 
each binding region at various levels of theory.
NG/G in brackets stands for 'No Good'/'Good', respectively.
Statistical errors in the DMC results are indicated in parentheses.
}
\label{table:ng_table}
\end{table*}
\par
For further possible validations of our DMC-PES, 
we would take the facts that 
(a) The PESs are consistent with those estimated by 
another reliable standard, CCSD(T), and 
(b) We can make a plausible comparison that 
explains the experimentally observed density 
from our estimated binding lengths $R_e$. 
For (a), we provide detailed discussions on 
the comparison as well with DFT later 
(see ``Calibration of DFT''). 

\par
As for (b), our scheme that relates 
$R_e$ with an experimental density 
is confirmed to work well not only for CHS but also 
for benzene molecules as described in Appendix E. 
For CHS, experimental values of 
the molecular weight ($180.61$ g/mol) 
and density ($0.97$ g/cm$^3$ at $T = 298.15$K) 
lead to the mean inter-molecular distance, $R_{\rho} = 6.8$ \AA{}, 
which fairly reasonably 
drops within the binding lengths of Type-A to C.
As shown in Table~\ref{table:two-body},
the simple thermal averaging over the three configurations
by the factor $p\sim \exp (-\Delta E/kT)$ gives us an 
underestimation $\bar{R}_{\rm dim}\sim 4.9$ \AA{} 
compared with $R_{\rho}$. 
An alternative averaging 
over the 'diagonal lengths' of four-body trapezoids, 
as shown in Fig.~\ref{tetramer-pattern},
gives an improved estimate, $\bar R_{\rm tetra}$,
getting closer to the experimental 
estimation of $6.8$\AA{}, as shown in Table~\ref{table:four-body}.
\begin{figure}[htb]
\begin{center}
\includegraphics[scale=0.38]{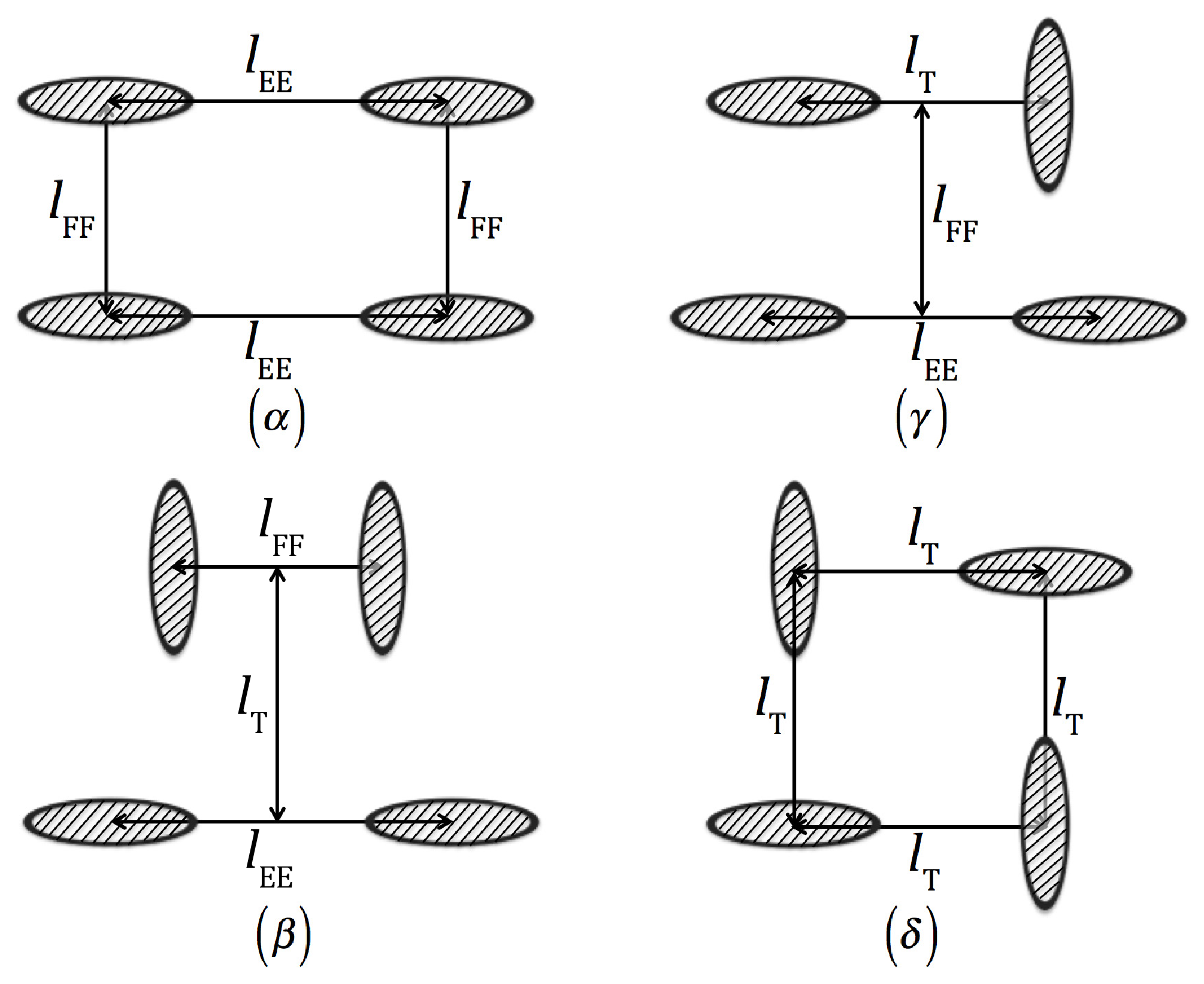}
\end{center}
\caption{
Possible four-body clusterings formed from 
the two-body coalescences shown in Fig.~\ref{dimer-pattern}.
Hatched regions stand for the surfaces 
surrounded by the ring of cyclohexasilane molecule.
$l_{\rm FF}$, $l_{T}$, and $l_{\rm EE}$ 
correspond to the binding lengths 
with 'face-to-face', 'T-shape', and 'edge-to-edge'
configuration, respectively.
}
\label{tetramer-pattern}
\end{figure}

\begin{table*}[htb]
\begin{tabular}{ccccc}
  &  B3LYP-GD3 & MP2 & CCSD(T) & DMC \\ [2pt]
\hline
$p(\mathrm{\alpha})$/$R_{\rm diag}$  & $0.343/6.6$ & $0.293/6.3$
& $0.327/6.4$ & $0.48(27)/6.5$ \\ [2pt]
$p(\mathrm{\beta})$/$R_{\rm diag}$   & $0.243/6.7$ & $0.248/6.3$
& $0.245/6.5$ & $0.21(7)/6.6$ \\ [2pt]
$p(\mathrm{\gamma})$/$R_{\rm diag}$  & $0.243/6.6$ & $0.248/6.1$
& $0.245/6.3$ & $0.21(7)/6.4$ \\ [2pt]
$p(\mathrm{\delta})$/$R_{\rm diag}$  & $0.172/6.6$ & $0.210/6.1$
& $0.183/6.2$ & $0.09(19)/6.4$ \\ [2pt]
$\bar{R}_{\rm tetra}$                & $6.6$       & $6.2$       &
$6.4$
& $6.5(2)$ \\ [2pt]
\hline
\end{tabular}
\caption{
Comparisons of the equilibrium stability among 
four clustering configurations in 
Fig.~\ref{tetramer-pattern}, in terms of the 
thermal probability weight, $p \sim \exp (-\Delta E/kT)$.
$R_\mathrm{diag}$ [\AA] stands for the diagonal length
for each tetramer, and
$\bar R_{\rm tetra} $~[\AA] is the thermal averaged 
diagonal length at $T = 298.15$ K.
}
\label{table:four-body}
\end{table*}

\subsection{Calibration of DFT}
DFT is a much more practical choice of methods combined with our
{\it ab initio} Hamaker evaluation schemes, but its reliability
strongly depends on XC functionals adopted as usual. 
Here we provide a useful calibration over several XC functionals 
appropriate for predicting $A$ values,
by comparing with the many-electron wavefunction theories.

\par
Fig.~\ref{PESsummary} highlights 
typical binding curves evaluated by various methods, though only for Type-A
(for the other types, see Supporting Information).
All the SCF curves were corrected by 
the BSSE scheme~\cite{1970BOY,1994DUI,1996SIM,2000HEL}
(see Supporting Information).
The present study takes CCSD(T) as a standard reference to
calibrate the performance of the SCF approaches.
We can find the DFT predictions scattering around CCSD(T).
Except LDA, conventional functionals such as PBE and B3LYP
fail to capture the binding itself.
The LDA overbinding has been frequently reported for
several molecular bindings.~\cite{2013HON,2010HON,2012WAT,2015HON,2008TKA}
This can be regarded as spurious due to improper self interactions: 
Exchange repulsion is not fully reproduced in LDA 
because of the lack of the exact cancellation of 
self interaction, and hence spurious 'chemical' bindings
are formed due to the weakened repulsions, rather than 
true molecular bindings.
The exchange repulsion weakened in LDA gets recovered 
when changing XC into GGA and further into B3LYP, 
which may explain the repulsive curves pushing 
the minimum toward distant region.
As LDA is known to inherently fail to describe
dispersion interactions, 
a significant difference in the LDA estimations between
$A_\mathrm{add}(C_6^\mathrm{stable;LOG})$ and 
$A_\mathrm{add}(C_6^\mathrm{stable;LJ})$ 
(see Table~\ref{table:physicalquantities})
implies a poor reliability on its long-range
behavior description.

\begin{figure}[htb]
\begin{center}
  \includegraphics[scale=0.85]{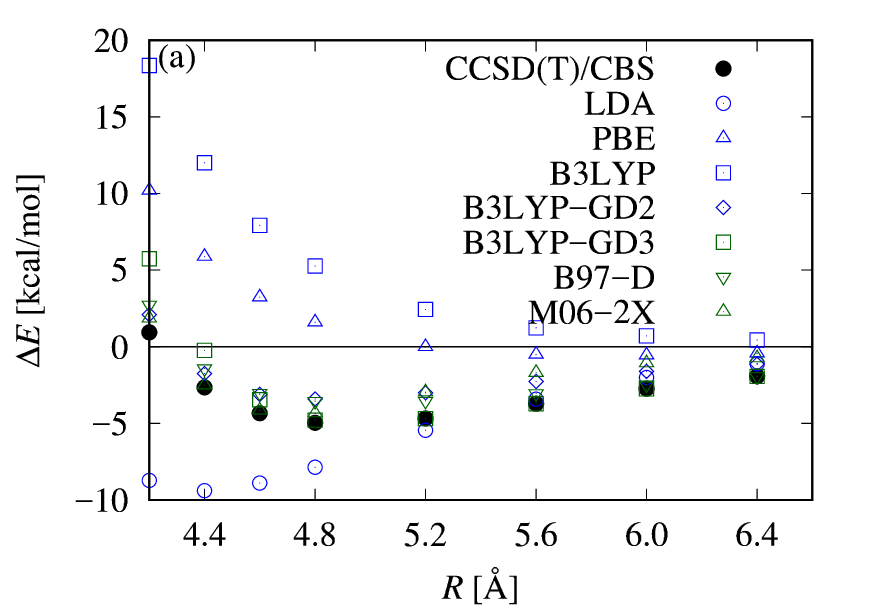}
  \includegraphics[scale=0.85]{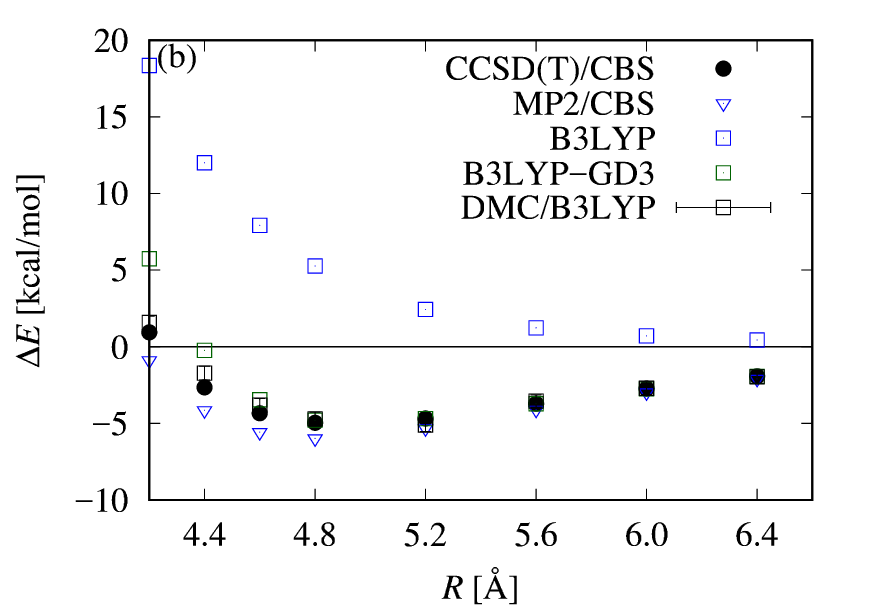}
\end{center}
\caption{
Binding curves of Type-A (parallel) dimer coalescence 
evaluated by (a) DFT methods (LDA, PBE, B3LYP, B3LYP-GD2/GD3, 
B97-D, M06-2X) and (b) correlation methods (MP2, CCSD(T) and 
DMC/B3LYP) compared with selected DFTs. 
All the curves (except DMC) are corrected by BSSE 
and CBS (see Supporting Information for more details).
}  
\label{PESsummary}
\end{figure}

\par
The XC functionals for molecular interactions, M06-2X, B97-D 
and B3LYP-GD(2,3), on the other hand, well reproduce 
the bindings at their equilibrium lengths, as seen in Fig.~\ref{PESsummary}.
We see, however, that M06-2X and B3LYP-GD2 
give rise to less reliable asymptotic behaviors at long-range region,
where they decay much faster than CCSD(T) or the other XC functionals for
molecular interactions.
As for M06-2X, its functional form based on hybrid meta-GGA does not
explicitly contain dispersion interactions by its construction,
and its parameterizations of the XC functionals 
are adjusted so as to reproduce a number of molecular bindings
{\it around} their equilibrium geometries,
giving rise to the unreliable long-range behavior.
B97-D and B3LYP-GD2 are classified into the DFT-D2 family including 
'atom-pairwise' second-order perturbative dispersion corrections
(two-body term).~\cite{2006GRI}
Both B97-D and B3LYP-GD2 give poor estimates of binding energies and lengths,
but the former behaves better than the latter at long-range region,
being appropriate for the estimation of Hamaker constants.
This implies long-range behaviors also depend on original functionals,
and atom-pairwise dispersion corrections do not necessarily lead to
a correct description of 'molecule-pairwise' dispersion interactions,
as in the B3LYP-GD2 case.
It has been reported that DFT-D3 including 
atom-pairwise third-order perturbative dispersion corrections (three-body term)
can remedy this kind of discrepancy in long-range as well as equilibrium
behaviors 
at the DFT-D2 level of theory{~\cite{2010GRI}}.
It is notable that the present B3LYP-GD3 binding curve is well improved 
in its long-range behavior to reproduce a correct decaying exponent.
For the present CHS case, its correct molecule-pairwise dispersion behavior
at long-range region
requires both the second- and third-order perturbative dispersion corrections.
Looking at the short-range region, on the other hand, 
we find that B97-D and M06-2X give a better description than B3LYP-GD3, 
getting closer to the DMC and CCSD(T) estimations.
This suggests that B3LYP-GD3 includes too large Hartree-Fock exchange effects
to be adequately canceled out by correlation effects.
The above results can be summarized in Table~\ref{table:ng_table}.

\par
Our DMC and MP2 results are shown in Fig.~\ref{PESsummary} (b),
compared with the reference CCSD(T), the typical SCF (B3LYP),
and the best within DFT at equilibrium and long-range regions (B3LYP-GD3).
As is well known, MP2 overbinds with deeper (shorter)
binding energy (distance).~\cite{2012RIL}
It may not be surprising to get the coincidence of
asymptotic behaviors between MP2 and CCSD(T), because 
the present CCSD(T) is corrected by the CBS scheme taken 
from MP2~\cite{2004SIN} (see Supporting Information).
Three DMC curves were obtained starting from guiding functions
generated by LDA, PBE, and B3LYP, respectively (see Supporting Information). 
They almost converged to the same binding curve, even starting 
from either B3LYP (worst in reproducing binding at SCF level) or
LDA (too deep spurious overbinding at SCF level).
Similar insensitivity to the choice of guiding functions has been also
reported for a DNA stacking case,{~\cite{2013HON}}
implying that these DMC predictions are not seriously affected by
the fixed-node approximation.
Based on the variational principle with respect to 
nodal surfaces in DMC{~\cite{1982REY,1991MIT,2006CAS}},
we henceforth concentrate on the B3LYP guiding function only, 
because it gives the lowest total energy though the energy differences
among the three binding curves are quite small.
Note that this is consistent with a number of previous DMC 
studies~\cite{2013HON,2012HON,2010KOL,2012MAN,2012HON2}

\par
The present DMC is found to give almost the same results as CCSD(T).
A remarkable difference between CCSD(T) and DMC is 
the binding energy at short range, $\Delta E(4.2)$ by 
$\sim 0.6 (\pm 0.4)$ kcal/mol.
The difference would be partly attributed to the 
dynamical correlation effect, which becomes more important 
at shorter binding length as well as exchange repulsions.
Even under the fixed-node approximation, the dynamical correlation
is expected to be well 
described,
~\cite{2010NEE,2012AUS,2008KOR,2013HOR,2014HOR,2013DUB,2014DUB,2013HON,2010HON,2012WAT,2015HON}
and hence the present DMC curve is regarded as
the best description of the binding of CHS.

\subsection{Practicality: DMC vs. CCSD(T)}
Fig.~\ref{dmc-ccsdt} shows the comparison between DMC and CCSD(T)
with and without basis set (CBS) corrections.
Even though CCSD(T) is known as the 'gold standard'
among {\it ab initio} predictions,
the practical use of CCSD(T) requires very careful 
handling of corrections, as described in Supporting Information, 
to get enough reliable predictions.~\cite{2013REZ}
The correction itself is also under quite a limited
approximation~\cite{1998TRU,2004SIN} (see Eq.~(1) in Supporting Information).
These practical limitations are, in contrast, 
not the case in DMC because it is free from
the basis set choice to the extent that only the
nodal structure of 
the many-body wavefunction is fixed by the given
basis set.
In order to evaluate Hamaker constants of practically larger systems, therefore,
DMC has the advantage over CCSD(T) with less sensitivity to basis sets.
\begin{figure}[htb]
\begin{center}
  \includegraphics[scale=0.85]{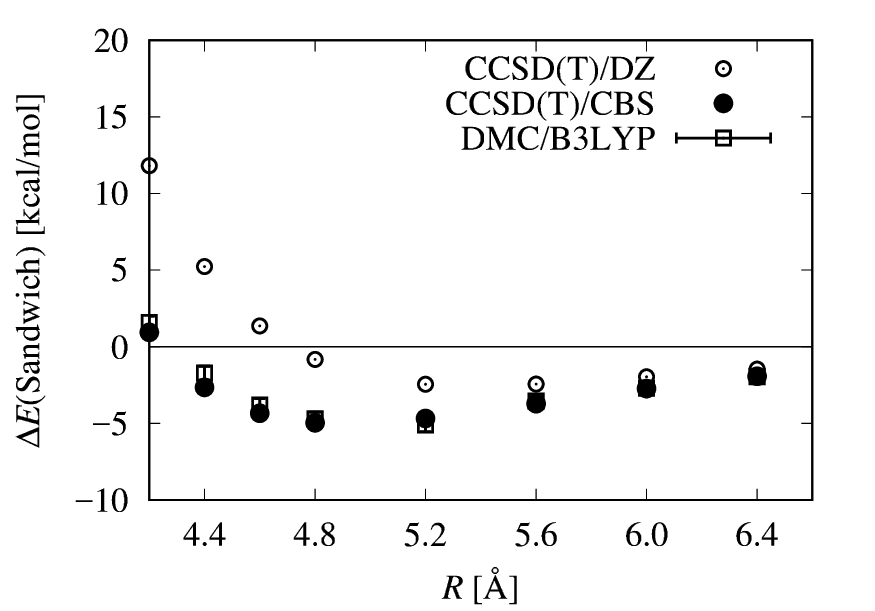}  
\end{center}
\caption{
  Comparison of binding curves between DMC and
  BSSE-corrected CCSD(T)
  with the CBS/DZ basis set.
'CCSD(T)/DZ[CBS]' stands for the raw value without any corrections
by DZ basis sets [CBS limit], while 'BSSE-CCSD(T)/DZ[CBS]' means
that with BSSE corrections.
}
\label{dmc-ccsdt}
\end{figure}

\section{Concluding Remarks}
\label{conclusion}
We considered a scheme using DMC-PES to evaluate Hamaker
constants $A$ 
for practical anisotropic molecules,and applied it to a 
cyclohexasilane (CHS) molecule used as an ink for printed electronics. 
The scheme should take into account two important factors 
for practical applications, namely the weak molecular 
interactions dominated by electron correlations (especially dispersion), and 
non-unique coalescing direction between anisotropic molecules. 
By making comparisons with the estimations by Lifschitz theory ($A_\mathrm{L}$) 
on benzene, we clarified several possible origins to give 
systematic biases on $A_{\rm add}$ when it is estimated 
by PES with/without any averaging operations over anisotropy. 
The success of our scheme in the benzene case leads us to its application to CHS.
In the application to CHS, 
our DMC results coincides fairly well with other 
correlation methods such as CCSD(T), MP2, and several 
DFT with exchange-correlation functionals for 
molecular interactions, like B3LYP-GD3. 
The evaluated binding curve can be reasonably 
validated by the experimentally observed 
density of the liquid solution via a scheme 
to relate its binding length and the mean 
inter-molecule distance. 
We find out that the parallel-wise coalescence of 
molecules gives the longest distant exponent 
for the interaction, being around 6.0. 
Several possible fitting schemes are applied 
to get $A_{\rm add}$, and finally we estimate it 
around $105 \pm 2$ [zJ], 
with practically enough small statistical error. 
Though there is no experimental data available for a direct
comparison, the present estimation is well supported
from the trend of both 
Hamaker constants for similar kinds of molecules 
and of systematic difference between the predictions
by the Lifshitz theory and by the asymptotic 
exponent estimations. 


\section{Associated Content}
\label{asscont}
The BSSE and CBS corrections to the SCF and correlated methods and 
the time-step bias in DMC are discussed in more detail at
Supporting Information. 
This material is available free of charge via the Internet at
\url{http://pubs.acs.org}

\section{Acknowledgments}
The authors thank Mr. M. Imamura for his preliminary calculations.
K.H. is grateful for financial support from a KAKENHI grant (15K21023), 
a Grant-in-Aid for Scientific Research on Innovative Areas (16H06439), 
PRESTO and the Materials research by Information Integration Initiative (MI$^2$I) project 
of the Support Program for Starting Up Innovation Hub 
from Japan Science and Technology Agency (JST).
R.M. is grateful for financial support from MEXT-KAKENHI 
grants 26287063 and that from the Asahi glass Foundation.
The computation in this work has been mostly done
using the facilities of the Center for Information Science in JAIST.

\section{Appendix}
\subsection{A. Computational Methods}
\label{methods}
The binding curves are evaluated by DMC, 
compared with CCSD(T), MP2, and several DFT
calculations with various XC functionals.
As a common choice, the fixed-node 
approximation{~\cite{2010NEE,2012AUS}} 
was made to the DMC simulations (DMC),
taking Slater-Jastrow wavefunctions as the guiding functions.
The Slater determinants are composed of
Kohn-Sham (KS) orbitals obtained using
Gaussian09{~\cite{2009FRI}} with
Burkatzki-Filippi-Dolg (BFD) 
pseudo potentials (PP){~\cite{2007BUR}}
and its accompanying VTZ Gaussian basis sets.
The BFD-PPs have been proved to give enough practical 
accuracies not only in DMC but also DFT on the 
applications such as a DNA stacking problem.{~\cite{2013HON}}

\par
Our Jastrow functions~\cite{1955JAS,2004DRU}
were those implemented in \texttt{CASINO}~\cite{2010NEE},
consisting of one-, two-, and three-body contributions,
denoted as $\chi$-, $u$-, and $F$-terms, respectively.
The $\chi$-, $u$-, and $F$-terms include 16, 16, and 32
adjustable parameters, respectively.
They were optimized by the variance minimization
scheme{~\cite{2005UMR,2005DRU}}.
The electron-electron cusp condition~\cite{1957KAT}
was imposed only on the $u$-term during the optimization procedure.
For DMC statistical accumulations, we set
the target population (the number of random walkers) to
be 1,024 configurations in average
and the time step to be $\delta t$ = 0.02 in atomic unit. 
The time step bias~\cite{1993UMR} arising from this choice 
is discussed in Supporting Information.
We took averages over $1.7\times10^5$ accumulation steps after 
the equilibration of $10^3$ steps.
We also used $T$-move scheme~\cite{2006CAS}
for the locality approximation to the evaluation of
PPs~\cite{1991MIT,1982REY} in DMC.

\par
Only for Type-A, we benchmarked various 
DFT-SCF calculations for a comparison with DMC, 
seeing how the choice of the XC functionals
affects the trial nodal structures in DMC. 
Our choice of XC functionals in DFT includes
those recently designed for molecular interactions,
B3LYP+GD2~\cite{2006GRI}/GD3,~\cite{2010GRI} 
M06-2X,~\cite{2008ZHA} and B97-D,~\cite{2006GRI}
as well as LDA~\cite{1980VOS}, PBE~\cite{1996PER},
B3LYP~\cite{1988LEE,1993BEC,1994STE}. 

For a systematic comparison, we consistently 
used the same basis sets as DMC, 
VTZ basis sets provided in BFD-PP library~\cite{2007BUR}.
For correlated methods (MP2 and CCSD(T)), however,
the VTZ is too large to be accommodated in 
tractable memory capacities 
(512GB shared by 64 parallel cores in SGI Altix UV1000).
To correct biases due to basis sets choices, 
we considered Complete Basis Set (CBS)
methods~\cite{1998TRU,2004SIN} with two different basis sets, and
counterpoise methods for basis set superposition error 
(BSSE).~\cite{1970BOY,1994DUI,1996SIM,2000HEL}
Detailed discussions about these corrections 
are given in Supporting Information.
All the DFT-SCF and correlated calculations 
were performed using \texttt{Gaussian09}.~\cite{2009FRI}

\par
As demonstrated in Supporting Information, all the three functionals 
give almost the same binding energy and equilibrium distance,
but B3LYP is found to give the best nodal surface
in the sense of the variational principle.
Hence we concentrated only on B3LYP 
orbitals for DMC for Type-B and -C.

\subsection{B. Summary of formalisms of Hamaker constants}
In most of practical cases, the Hamaker constants are estimated 
by the macroscopic frameworks based on Lifshitz theory~\cite{1956LIF} 
(let us denote the estimation by this frameworks as $A_\mathrm{L}$). 
In the scheme, $A_\mathrm{add} = \pi C_6 \rho^2$, 
several possibilities are available for $C_6$ 
evaluations, including those 
(i) by DOSD (dipole oscillator strength distribution)
experiments,~\cite{1992KUM}
(ii) by estimations by the Casimir-Polder relation
(CPR)~\cite{1948CAS} using 
{\it ab initio} evaluations of dynamical
polarizabilities,~\cite{2005JIE,2007MAR}
and 
(iii) by the fitting of asymptotic behaviors 
in the molecular binding curves, or potential 
energy surfaces (PES), evaluated by 
{\it ab initio} calculations.~\cite{2006POD}
The Casimir-Polder formula for (ii) is given in hartree units as,
\begin{equation}
  \label{cpr}
  C_6 = \frac{3}{\pi}\int_0^\infty du \cdot \bar{\alpha}(iu)^2 \ ,
\end{equation}
in an integral over the imaginary frequency, $iu$, 
of the orientation average of the polarization tensor, 
$\bar{\alpha}(iu): = (1/3)\mathrm{Tr}\left[ { \alpha}(iu)\right]$. 
$\bar{\alpha}$ can be evaluated by TDDFT (time-dependent DFT) 
within the linear response theory.~\cite{2005JIE,2007MAR}
Provided that the molecule has a unique 
absorption frequency (ionization energy), $\nu_I (= I/h)$, 
a further approximation with 
$\bar{\alpha}(iu) \approx \bar{\alpha}(0){\nu_I^2}/({u^2+\nu_I^2})$ 
substituted to (eq.~\eqref{cpr}) leads to the London formula
of the dispersion force~\cite{1937LON}, 
\begin{equation}
  \label{london}
  C_6 = \frac{3}{4}{\bar{\alpha}^2(0)}I \ ,
\end{equation}
where $\bar{\alpha}(0)$ is the static polarizability. 

\par
$C_6$ for practical anisotropic molecules obviously 
depends on the orientation of coalescence, such as T-shape, 
parallel, sandwich, {\it etc}. 
Plausible averaging is required over the 
possible orientations to get $A_\mathrm{add}$, 
which is the main subject of the present study. 
This would be a reason that $A_\mathrm{L}$ is used 
much rather than $A_\mathrm{add}$ because 
in the former the non-additivity as well as 
the anisotropy are effectively taken into 
account by using macroscopically averaged 
quantities. 
In (i) and (ii), the macroscopic/observed quantities 
used in the formula would be regarded as 
the effective consideration of such averaging to give 
$\langle C_6\rangle$, as we used in Table~\ref{table:benzene1}. 
For most of the practical cases, the Hamaker constants are 
evaluated not by $A_\mathrm{add}$ but by $A_\mathrm{L}$, 
a macroscopic framework based on 
Dzyaloshinskii-Lifshitz-Pitaevskii(DLP) theory, 
in which the Hamaker constant is expressed as an 
infinite series of an expansion. 
Truncation upto the second term gives a practical 
approximation, known as Ninhan-Parsegian 
formula~\cite{1969PAR,1970NIN}, 
\begin{equation}
  A_{L} = \frac{3}{4}kT\left(\frac{\varepsilon - 1}
  {\epsilon + 1} \right)^2 + \frac{3 h \nu_e}{16\sqrt{2}}\cdot
  \frac{(n^2-1)^2}{(n^2 + 1)^{3/2}} \ ,
\end{equation}
and its truncation error is estimated to be
around 5\%.~\cite{2011ISR} 
The Hamaker constant can then be evaluated 
using macroscopic quantities of the bulk, {\it i.e.}, 
dielectric constant $\varepsilon$, and 
the refractive index $n$ 
($k$ and $T$ are the Boltsmann constant and absolute 
temperature, respectively, while $h$ and $\nu$ are 
the Planck constant and the frequency of the primary 
electronic excitation in ultra-violet range). 
Unlike $A_\mathrm{add}$, the macroscopic $A_{L}$ can 
avoid the additive assumptions, namely 
the macroscopic quantities effectively take into 
account the non-additivity as well as the anisotropy.

\subsection{C. Comparison of different predictions of $A$}
For the side-way manner of the validation 
of $A$ predicted for CHS, we would like to 
know if there is a systematic bias between 
$A_\mathrm{add}$ and other macroscopic $A_\mathrm{L}$.
For this purpose,
we take benzene as a representative tiny benchmark.
In this case, there are many references to $C_6$ and $A_\mathrm{L}$ available
in literature~\cite{2011ISR,2005JIE,2007MAR,1992KUM,2006POD},
by which we can survey the possible 
relation between $A_\mathrm{add}$ and $A_\mathrm{L}$ 
to get a plausible validation for the estimate of 
realistic Hamaker constants $A_\mathrm{add}(C_6)$
evaluated from {\it ab initio} PES calculations.
Even for this simple molecule, 
there has been little investigations 
relating it to $A_\mathrm{add}$ for practical molecules, 
though it is straightforward. 
This might be attributed to the difficulty of 
the averaging over anisotropic configurations of coalescence. 

\begin{table*}[htb]
\begin{tabular}{llcl}
  Label & Scheme/Theory/Method & $C_6$/$10^3$ [au] & $A$ [zJ] \\
\hline
1/$A_\mathrm{L}$  & Exp/DLP      & NA
& $50 \pm 2$\footnotemark[1] \\
\hline
2/$A_\mathrm{add}(\langle C_6^\mathrm{London} \rangle^{\rm iso})$
 & London/HF & $0.925$\footnotemark[2] & $42$\\
3/$A_\mathrm{add}(\langle C_6^\mathrm{London} \rangle^{\rm iso})$
 & London/B3LYP & $1.105$\footnotemark[3] & $48$\\
4/$A_\mathrm{add}(\langle C_6^\mathrm{CPR} \rangle^{\rm iso})$
 & CPR/TDHF & $1.737$\footnotemark[4] & $75$\\
5/$A_\mathrm{add}(\langle C_6^\mathrm{CPR} \rangle^{\rm iso})$
 & CPR/TDDFT & $1.773$\footnotemark[5] & $77$\\
6/$A_\mathrm{add}(\langle C_6^\mathrm{DOSD} \rangle^{\rm iso})$
 & Exp/DOSD     & $1.723$\footnotemark[6] & $74$ \\
\hline
7/$A_\mathrm{add}(C_6^\mathrm{PES})$(Sandwich)  & PES/MP2      
 & $0.59$\footnotemark[7] & $25$ \\
8/$A_\mathrm{add}(C_6^\mathrm{PES})$(Sandwich)  & PES/CCSD(T)  
 & $0.602$\footnotemark[8] & $26$ \\
9/$A_\mathrm{add}(C_6^\mathrm{PES})$(T-shape)  & PES/CCSD(T)  
 & $3.911$\footnotemark[9] & $169$ \\
\hline
10/$A_\mathrm{add}(\langle C_6^\mathrm{PES}\rangle^{\rm iso})$
 & PES/DFT-SAPT & $1.726$\footnotemark[10] & $74$ \\
11/$A_\mathrm{add}(\langle C_6^\mathrm{PES}\rangle^{\rm iso+aniso})$
 & PES/DFT-SAPT & $1.165$\footnotemark[11] & $50$ \\
\hline
12/$A_\mathrm{add}(C_6^\mathrm{PES})$(ParaDisp) &PES/DMC/log  
 & $1.25 \pm 0.27$\footnotemark[12] & $54 \pm 12$ \\
13/$A_\mathrm{add}(C_6^\mathrm{PES})$(ParaDisp) &PES/DMC/LJ   
 & $1.19 \pm 0.10$\footnotemark[13] & $51 \pm 4$ \\
 \hline
\end{tabular}
\caption{
  Comparisons of Hamaker constants $A$ of Benzene estimated by
  different schemes. The braket, $\langle\cdots\rangle$,
  means spatial averaging (see text for details). 
}
\label{table:benzene1}
\end{table*}
\footnotetext[1]{Ref.~\cite{2011ISR}}  
\footnotetext[2]{Ionization energy and static polarizability were
  obtained at HF/6-311++G(3d,3p) level.}
\footnotetext[3]{Ionization energy and static polarizability were
  obtained at B3LYP/cc-pVQZ level.}
\footnotetext[4]{Ref.~\cite{2005JIE}} 
\footnotetext[5]{Ref.~\cite{2007MAR}} 
\footnotetext[6]{Ref.~\cite{1992KUM}}         
\footnotetext[7]{Ref.~\cite{2006ZEI}}  
\footnotetext[8]{Ref.~\cite{2004SIN}} 
\footnotetext[9]{Ref.~\cite{2004SIN}} 
\footnotetext[10]{Ref.~\cite{2006POD}} 
\footnotetext[11]{This work with data from Ref.~\cite{2006POD}.}
\footnotetext[12]{This work with data from Ref.~\cite{2015AZA}}
\footnotetext[13]{This work with data from Ref.~\cite{2015AZA}}

\par
Table~\ref{table:benzene1} summarizes the Hamaker constants 
estimated by several approaches explained in Appendix B.
No.2-6 in Table~\ref{table:benzene1} are obtained using $C_6$ 
evaluated by London's theory, CPR, and DOSD, 
which can be regarded as an equivalent averaging 
over orientations to get a representative 
isotropic value, $\langle C_6\rangle^{\rm iso}$. 
For London's theory (No.2/3), the closer values to 
$A_\mathrm{L}$ (No. 1) might be accidental. 
These should be comparable rather to No.4/5 but turned 
out to be underestimated by around 40\%{}. 
The underestimation can be explained because, as we mentioned, 
the London theory only picks up single absorption frequency, 
ignoring other contributions which are all positive. 
No.4-6 are consistent with each other being around 75 [zJ], 
but overestimating when we take $A_\mathrm{L}$ as the reliable reference 
for the perfect averaging about the anisotropy and non-additivity. 
The importance of the anisotropy can be seen in No.7-9, 
where $C_6$ are evaluated only by PES for a coalescence 
configuration, such as Sandwich or T-shape. 
We see that different methods for PES give consistent 
results with each other for the same configuration (No.7 and No.8), 
while the same method gives the different estimation 
for different configurations (No.8 and No.9). 

\par
SAPT evaluations, No.10 and No.11, give
some confidence about the effective isotropic averaging 
for No.4-6 and the importance to consider the anisotropy 
for the $A_\mathrm{L}$ value. 
In SAPT, the dispersion interaction is evaluated in the form,  
\begin{equation}
  \begin{split}
  E_\mathrm{disp} \sim \frac{1}{R^6} \cdot & \sum_{l_A,k_A,l_B,k_B,l}
  C_6[l_A,k_A,l_B,k_B,l] \\ & \cdot w_6[l_A,k_A,l_B,k_B,l]
  (\omega_A,\omega_B,\Omega) \ ,
  \end{split}
\end{equation}
as the summation over the possible anisotropic configurations 
labelled by the rank of tensor $\{l_A,k_A,l_B,k_B,l\}$ with 
each weight $w_6$. 
$\{\omega_A,\omega_B,\Omega\}$ denote the Euler angles 
between molecules A and B, and the solid angle between the molecules, 
respectively. 
No.10 is evaluated from the isotropic contribution, 
$C_6[0,0,0,0,0]=1726$, and consistent with No.4-6 as expected. 
We can also obtain the anisotropic contributions from 
the supplemental information of the paper~\cite{2006POD}, 
$C[0,0,2,0,2] = C[2,0,0,0,2] = -552, C[2,0,2,0,0] = 17,
C[2,0,2,0,2] = 45, C[2,0,2,0,4] = 482$, to estimate No.11.
To avoid the complicated averaging operations with serious 
weightings, we simply takes the arithmetic mean for 
isotropic and anisotropic contribution, 
and get $\langle C_6\rangle^{\rm iso+aniso}$ quite closer to $A_\mathrm{L}$ in No.1. 

\par
No.12/13 give the estimation by a single configuration, 
parallel-displacement (ParaDisp), which is identified 
as the most stable binding.~\cite{2006CHO}
The estimation is made from the DMC data 
by Azadi {\it et al.},~\cite{2015AZA}
from which we fit $C_6$ using log plots (No.12) 
or 6-12 Lennard-Jones (LJ) potential (No.13) 
as shown in Fig.~\ref{PESbenzene}. 
Since we are interested in the long-range 
asymptotic behavior $\sim R^{-6}$, we did not take 
the original spline-like fitting function 
used in the paper,~\cite{2015AZA} 
which is used to describe the whole PES shape 
being different from the present purpose. 
We note for the log plot that the larger errorbar
is a sort of inevitable consequence 
of log-plot for long-range exponents:~{\cite{2014MIS}}
For a fixed magnitude of statistical errors over 
the range of distance $R$, 
any decreasing dependence on a log-plot gives 
inevitably enlarged errorbars as $R$ increases
(the resolution of the vertical axis gets enlarged
downward by {\it definition}). 
In the present case, the statistical noise has 
been well suppressed less than $0.02$ kcal/mol, and 
further reduction of the errorbar in $A$ is
not practical. 
In LJ fitting, on the other hand, enough practical 
reduction of the errorbar has been achieved. 
The estimation of the fitting actually 
depends on the choice of the details of fitting functions, 
and data range of the fitting, for which we have chose it 
carefully with some validation as shown in Appendix D.
Despite a single configuration, 
the fitting for the parallel-displacement 
configuration, No.12/13, 
coincides well with $A_\mathrm{L}$. 
This implies that the most stable 
binding configuration 
(parallel displacement in this case) 
is almost dominant and other possible 
configurations can be ignored for $A$. 

\begin{figure}[htb]
\begin{center}
  \includegraphics[scale=0.85]{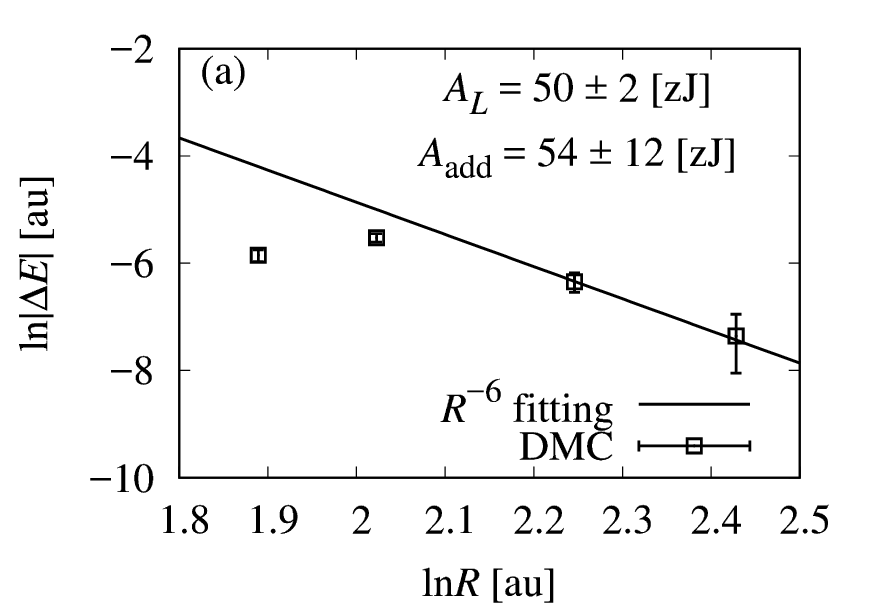}
  \includegraphics[scale=0.85]{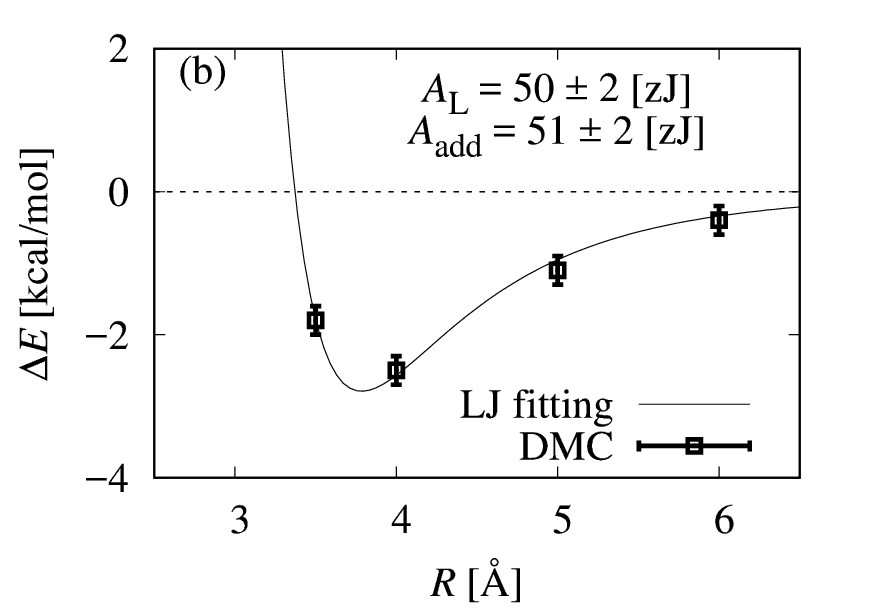}
\end{center}
\caption{
  DMC binding curves of benzene dimer:
  (a) logarithmic plot fitted by asymptotic $R^{-6}$ behavior
  and (b) Lennerd-Jones fitting.
}  
\label{PESbenzene}
\end{figure}

\subsection{D. Fitting scheme for $C_6$}
Several different fitting schemes are 
possible to extract $C_6$ from PES, in principle, 
such as log-fit ($C_6^{\rm stable;LOG}$), 
LJ-fit ($C_6^{\rm stable;LJ}$), 
and the power-fit for the 
correlation contribution ($C_6^{\rm stable;Corr}$). 
Table~\ref{table:benzene2} summarizes the fitting results 
of benzene using various kinds of fitting functions and 
the choices of data range to be fit. 
While every fitting seems to work fairly well 
as shown in Fig.~\ref{Fitbenzene}, 
the estimations of $A$ 
significantly depend on the arbitrary choice. 
We tried 6-12 LJ, 6-9 LJ, 6-exponential potential 
(6-exp)~\cite{1938BUC}, 
and pairwise polynominal fitting function.~\cite{2015AZA}
The choice of the fitting range is about whether 
we include the data at repulsive region (at $R=3.0$) or not. 
Only for 6-exp we could not get reasonable convergence without 
the data point at $R=3.0$ to increase the data points 
for such a strong non-linear fitting. 
For the polynominal, we could not extract $C_6$ for the 
asymptotic $R^{-6}$ behavior as we mentioned in the 
previous paragraph, but we can use it to get 
a reliable reference for the binding energy $\Delta E$ 
and bonding length $R_e$, as it is the most 
precise function for the whole fitting purpose as 
described in the paper.~\cite{2015AZA} 
Taking those reference for $\Delta E$ and $R_e$, 
we see that excluding the repulsion point, $R=3.0$, 
from LJ fitting gives better estimations. 
Though the log-fitting result (No.12) in 
Table~\ref{table:benzene1} has the large errorbar, 
the value is reliable to some extent in the aspect of
the asymptotic behavior, from which 
6-9 LJ (3.5 $\sim$ 6.0) gives larger deviation. 
Based on these facts, we finally take 6-12 LJ (3.5 $\sim$ 6.0) 
to provide the value (No.11) in Table~\ref{table:benzene1}. 

\begin{figure}[htb]
\begin{center}
 \includegraphics[scale=0.85]{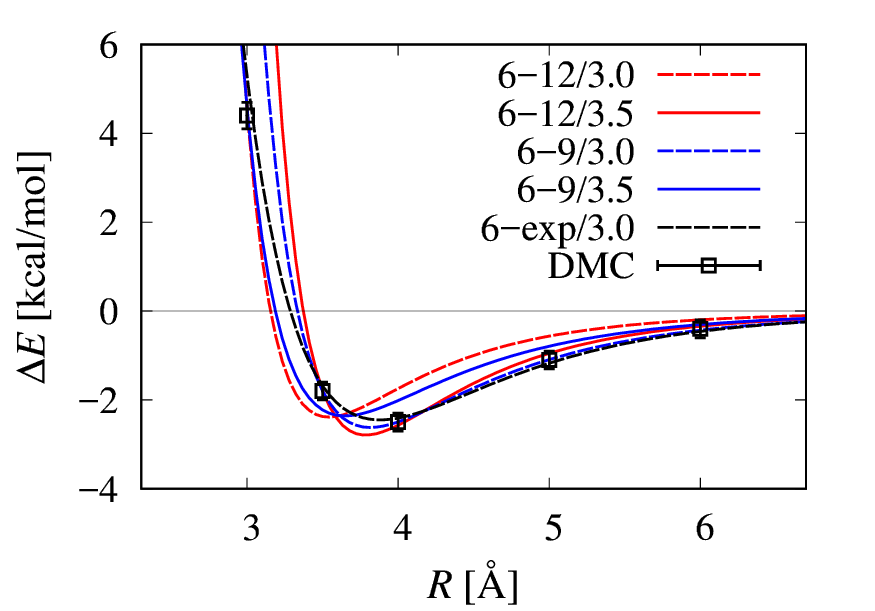}  
\end{center}
\caption{
Comparison between different fitting schemes of DMC 
binding curves 
(benzene dimer with parallel-displacement (PD) 
configuration) using 6-12 LJ (Lennard-Jones), 
6-9 LJ, and 6-exp with different fitting ranges 
({\it e.g.}, '6-12/3.0' means the range 
starting from $R$=3.0 [\AA{}]). 
}  
\label{Fitbenzene}
\end{figure}

\begin{table*}[htb]
  \begin{tabular}{ccccc}
    Functions & Range & $\Delta E$ [kcal/mol] & $R_e$ [\AA{}] & $A_\mathrm{add}$ [zJ] \\
    \hline
    6-12 LJ & $3.0\sim 6.0$ & $-2.4 \pm 0.2$ & $3.54 \pm 0.01$ & $29 \pm 2$ \\
    6-12 LJ & $3.5\sim 6.0$ & $-2.8 \pm 0.2$ & $3.78 \pm 0.02$ & $51 \pm 4$ \\
    6-9 LJ  & $3.0\sim 6.0$ & $-2.4 \pm 0.2$ & $3.64 \pm 0.01$ & $52 \pm 3$ \\
    6-9 LJ  & $3.5\sim 6.0$ & $-2.6 \pm 0.2$ & $3.81 \pm 0.03$ & $76 \pm 7$ \\
    6-exp   & $3.0\sim 6.0$ & $-2.5$ & $3.88$ & $70 \pm 11$ \\
    Pairwise Poly.~\cite{2015AZA}   & $3.0\sim 6.0$ & $-2.7 \pm 0.3$
    & $3.8 \pm 0.3$ & NA \\                
    \hline
    \hline    
    CCSD(T) reference\footnotemark[1] & Configuration & $\Delta E$ & $R_e$ & \\
    \hline    
    & PD (most stable) & $-2.78$ & $3.87$ & \\
    & T-shape  & $-2.74$ & $5.01$ & \\
    & Sandwich & $-1.81$ & $6.09$ & \\
    \hline
\end{tabular}
  \caption{
    Dependences of estimated binding energies ($\Delta E$), 
binding lengths ($R_e$), and Hamaker constants 
($A_{\rm add}$) on the choices of fitting functions 
and fitting ranges. 
Reference values for $\Delta E$ and $R_e$ 
by CCSD(T) are also shown.
}
\label{table:benzene2}
\end{table*}
\footnotetext[1]{CCSD(T) reference values obtained
  from Ref.~\cite{2004SIN}.}

\par
For CHS, the results using different fitting schemes
were tabulated in Table~\ref{table:physicalquantities} of main text.
We start with the log-fit ($C_6^{\rm stable;LOG}$).
To obtain plausible estimates of $C_6^{\rm stable;LOG}$,
it is essential to choose their fitting region $R_f$ at long-range.
We focus on the relation that by definition the binding energy
can be decomposed as the sum of 
Hartree-Fock (HF) and correlation contributions:
$\Delta E(R) = \Delta E_\mathrm{HF}(R) + \Delta E_\mathrm{corr}(R)$~\cite{1996HOB}.
Since $\Delta E_\mathrm{HF}(R)$ (exponentially) decays much faster 
than $\Delta E_\mathrm{corr}(R)$ (polynomially) at large $R$,
the asymptotic behavior is dominated by $\Delta E_\mathrm{corr}(R)$.
Thus, we choose $R_f$ such that $R \in R_f$ satisfies
$\left|E_\mathrm{HF}(R)/E_\mathrm{corr}(R)\right| < (1/10)$.
Asymptotic exponents can be extracted from the 
log-plots, as shown in Fig.~\ref{cyclo}.
The best fits of the exponents in DMC [CCSD(T)] give
$-5.6 \pm 5.8$ [$-5.9$], $-4.2 \pm 5.9$ [$-7.3$], and 
$-7.2 \pm 7.2$ [$-8.7$] for Type-A, B, and C, respectively.
Supported by the CCSD(T) estimations, 
we can somehow identify that Type-A dominates
the wettability with the longest-ranged exponent
that is almost close to the $\sim C_6/R^6$ dependence.
For Type-A, we can then identify the $C_6$ constant from 
the fitting with exponent fixed to be 6.0.
To sum up, only the stable configuration contributes
to $\sim 1/R^6$ asymptotic behavior 
while the other give different exponents. 
Note that this 'less contributions to $A$' from other 
meta-stable configurations is quite in contrast 
to the case for the molecular density estimation, 
for which only the stable binding configuration 
cannot reproduce the proper density, as describe 
in Appendix E.
\begin{figure}[htb]
\begin{center}
  \includegraphics[scale=0.85]{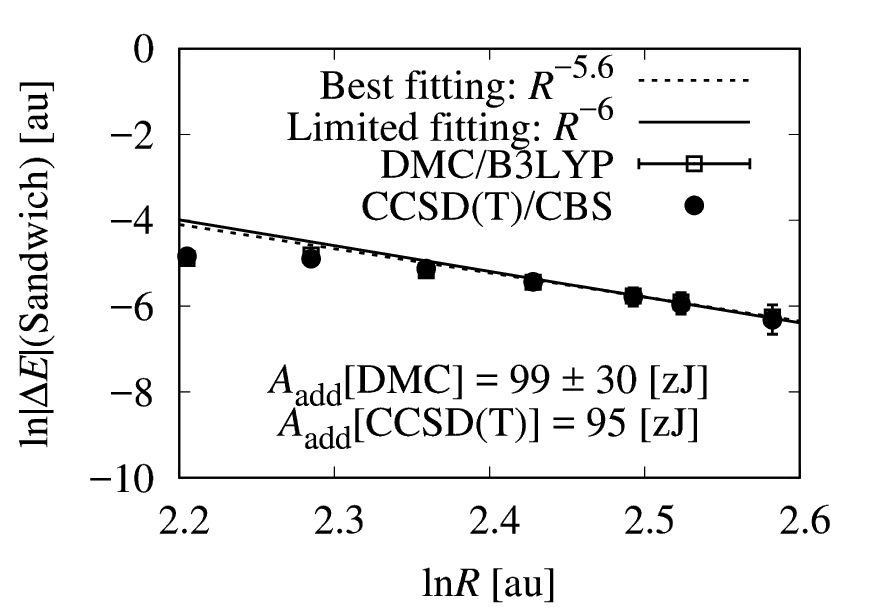}
  \includegraphics[scale=0.85]{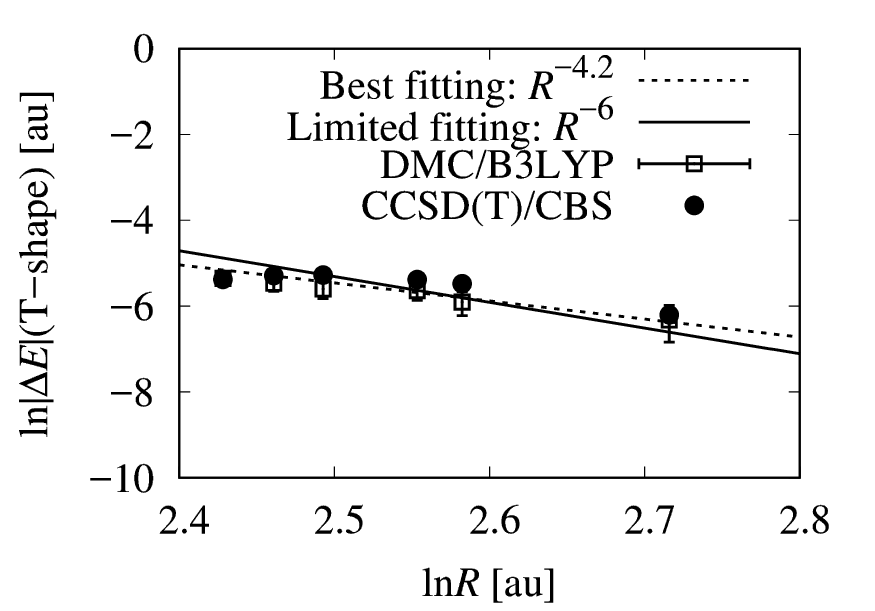}
  \includegraphics[scale=0.85]{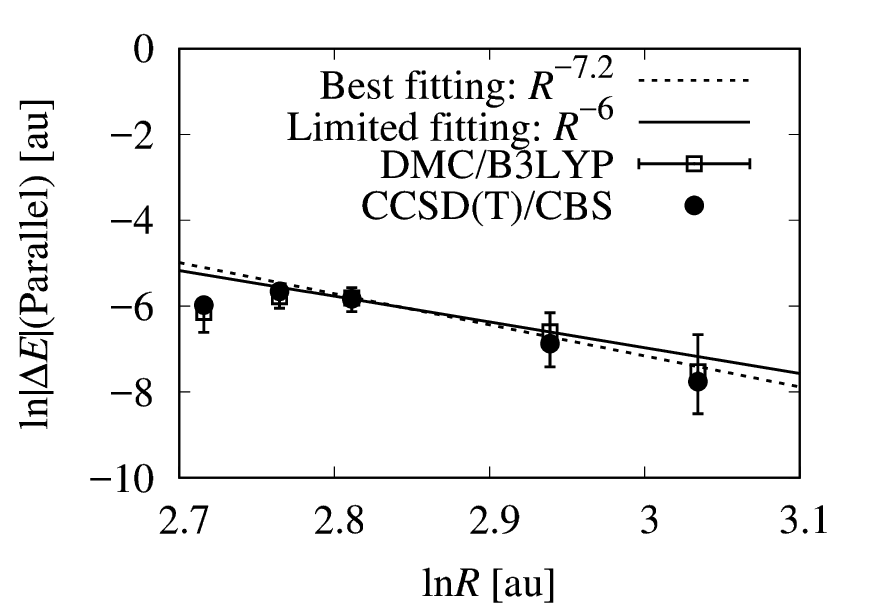}
\end{center}
\caption{Asymptotic behaviors of binding curves evaluated by 
DMC and CCSD(T), as given in logarithmic plots 
fitted by two different lines, 'Best' and 'Limited'.
In the former fitting, the exponent is fitted to 
get the best fitting while in the latter it is 
fixed to be assumed $R^{-6}$ behavior.}
\label{cyclo}
\end{figure}

\par
As was explained in the benzene case, 
the log-fits are inevitably accompanied by the larger 
statistical errorbars. 
If we aimed to reduce the errorbar by one more digit,
100 times more statistical accumulation would be necessary.
This computation corresponds to $2.2\times 10^6$ core-hours 
(half a year CPU time on 512 cores parallel, provided 
that we can keep on using it without any queue),
and hence is impractical.

\par
\begin{figure}[htb]
\begin{center}
   \includegraphics[scale=0.85]{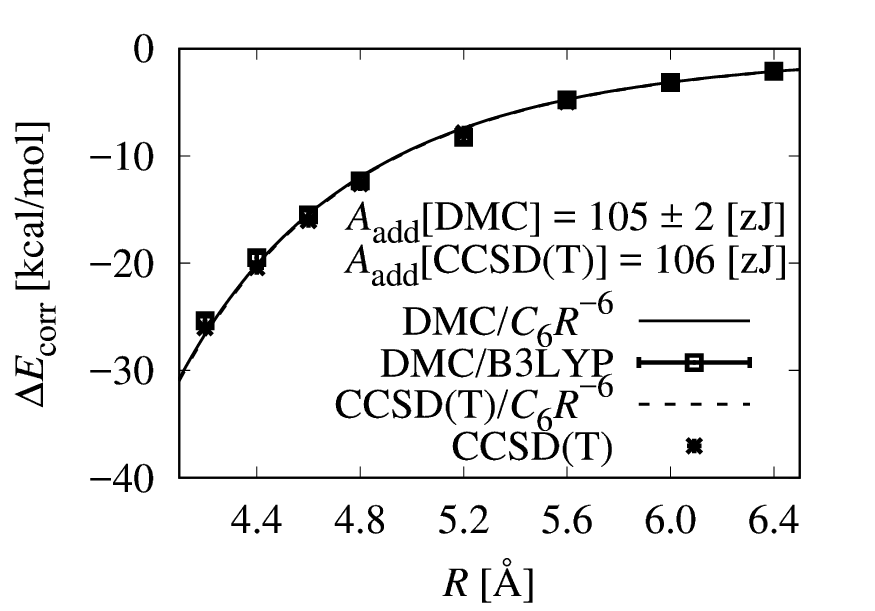}
\end{center}
\caption{
Correlation energy contributions to 
binding energies evaluated by DMC and CCSD(T).
}
\label{corr}
\end{figure}
The larger errorbars in the log-fitting are 
much improved by using the other fitting schemes:
The present study employed
the 6-12 LJ fitting (Fig.~\ref{ljfitting} of main text) using
\begin{equation}
  \label{lj1}
U(R) = -\frac{C_6}{R^6} + \frac{C_{12}}{R^{12}}.
 \end{equation}
and the correlation fitting (Fig.~\ref{corr}) based on
a power expansion,~\cite{1937LON,2016GRI} 
\begin{equation}
  \label{disp}
  \Delta E_\mathrm{corr}(R) \sim -\frac{C_6}{R^6},
 \end{equation}
to extract $C_6$ from PES.
As was discussed in the benzene case, 
the estimation depends on the data fitting range, 
and then some plausible choice is required. 
Table~\ref{table:chs2} compares the choices 
especially about whether the range includes 
repulsive region ($R < 4.4$) or not. 
To get the best choice, we adopted a figure of merit,
\begin{equation}
  \label{fm}
  f = \sum_{j=1} \frac{(\Delta E(R_j) - U(R_j))^2}{\sigma(R_j)^2} \ ,
 \end{equation}
as defined using the deviation of $\Delta E(R_j)$ from the fitting, 
weighted by the statistical error $\sigma(R_j)$ in DMC 
[which is set to be unity for CCSD(T)]. 
For the 6-12 LJ fitting, 
we chose the estimations achieving minimum $f$, those with 
the range $R = 4.6\sim 6.4$\AA{}, as finally tabulated in 
Table~\ref{table:physicalquantities}. 

\begin{table*}[htb]
  \begin{tabular}{ccccccc}
    Method & $U(R)$ & $R_f$ & $\Delta E$ & $R_e$ 
    & $A_\mathrm{add}$ & $f$\\    
    \hline
DMC & 6-12 LJ & $4.2\sim 6.4$ & $-4.9 \pm 0.2$ & $4.74 \pm 0.01$ 
& $85 \pm 4$ & $1.198$ \\
 & 6-12 LJ & $4.4\sim 6.4$ & $-5.2 \pm 0.2$ & $4.79 \pm 0.01$ 
& $100 \pm 5$ & $0.113$ \\
 & 6-12 LJ & $4.6\sim 6.4$ & $-5.3 \pm 0.2$ & $4.89 \pm 0.02$ 
& $107 \pm 7$ & $0.033$ \\
 & 6-12 LJ & $4.8\sim 6.4$ & $-5.2 \pm 0.2$ & $4.96 \pm 0.03$ 
& $112 \pm 9$ & $0.206$ \\
 & Corr. & $4.2\sim 6.4$ &  & 
& $103 \pm 1$ & $0.022$ \\
 & Corr. & $4.4\sim 6.4$ &  &
& $105 \pm 2$ & $0.005$ \\
 & Corr. & $4.6\sim 6.4$ &  &
& $108 \pm 2$ & $0.020$ \\
 & Corr. & $4.8\sim 6.4$ &  &
& $110 \pm 5$ & $0.062$ \\
 \hline
CCSD(T) & 6-12 LJ & $4.2\sim 6.4$ & $-5.10 $ & $4.76 $ 
& $86 $ & $ 0.027$ \\
 & 6-12 LJ & $4.4\sim 6.4$ & $-5.27 $ & $4.82 $ 
& $95 $ & $ 0.004$ \\
 & 6-12 LJ & $4.6\sim 6.4$ & $-5.24 $ & $4.89 $ 
& $103 $ & $ 0.002$ \\
 & 6-12 LJ & $4.8\sim 6.4$ & $-5.12 $ & $4.94 $ 
& $107 $ & $ 0.006$ \\
 & Corr. & $4.2\sim 6.4$ &  & 
& $ 106$ & $0.018 $ \\
 & Corr. & $4.4\sim 6.4$ &  &
& $ 108$ & $ 0.034$ \\
 & Corr. & $4.6\sim 6.4$ &  &
& $ 110$ & $ 0.053$ \\
 & Corr. & $4.8\sim 6.4$ &  &
& $ 110$ & $ 0.050$ \\
\hline
\end{tabular}
\caption{
  Dependence of the Hamaker constant $A_\mathrm{add}$ on 
  types of fitting functions $U(R)$ fitted within ranges $U_f$
  for DMC and CCSD(T). 
  The best choice of ranges were chosen such that
  values of figure of merit $f$ (see text for definition)
  achieves minimum.
  For the 6-12 LJ fitting,
  binding energies $\Delta E$ and binding lengths $R_e$
  were also listed.
}
\label{table:chs2}
\end{table*}

\par
For the correlation fitting, 
the final choices in Table~\ref{table:physicalquantities} of main text
are also those which achieve minimum $f$ in Table~\ref{table:chs2}, 
as we did for the 6-12LJ case.
The correlation fitting, if tractable, would be 
more plausible in the following sense:
(i) Its theoretical background is sound 
(perturbation theory on electron correlation at long-range).
(ii) Base on the theory, it is obvious to exclude
the repulsive (short-range) region from the fitting region.
(iii) Hence there is no ambiguity about
the model function to describe the repulsive region
such as 6-12/6-9 LJ or 6-exp.
(iv) Since $\Delta E_\mathrm{corr}$ increases monotonically, 
a better and more (numerically) stable fitting can be expected.

\subsection{E. Binding length and Density}
\label{de}

A PES gives a binding length $R_{e}$, which 
would have some relation to the experimental 
molecular density. 
Once some reliable relation was established, we can use 
it to validate the binding curve calculation. 
The relation is however not so clearcut as we describe below. 
The experimental density of the benzene liquid~\cite{2013HAY} gives an 
estimate of mean intermolecular distance as 
$R_{\rho} \sim 5.3$~\AA{}, being far larger than $R_{e}$ in 
the most stable binding by $26$\%. 
This is quite in contrast to the case of $A$ where 
only the most stable configuration seems dominant.
The simplest idea is to take into account the contributions 
not only from the most stable parallel displacement 
($\Delta E = -2.78$kcal/mol; $R_e = 3.87$\AA{}), but also 
other meta-stable ones, T-shape ($-2.74$kcal/mol; $5.01$\AA{}), 
and Sandwich ($-1.81$kcal/mol; $6.09$\AA{}). 
Only the most unstable configuration (Sandwich) has
a longer binding length of $R = 6.09$\AA{},
and the thermal averaging 
with the weight $p\sim \exp(-\Delta E/kT)$ at $T = 298.15$K 
gives $\bar R_{\rm dim} = 4.5$\AA{}, being 
$15$\%{} underestimation.

\par
As one of the possible origin for the discrepancy, 
we might consider the intra-molecular relaxation, but 
it is unlikely to account for it: 
the relaxation will bring energy gains at 
shorter binding lengths when the molecule deforms 
by the binding interaction, 
and hence make the binding length shorter, being further 
away from $R_{\rho}$. 

\par
Further consideration makes us realize that
we took into account
only two-body coalescences to argue the mean separation.
When we consider further four-body clusterings 
possibly occurring in realistic liquids, 
we notice that the mean separation seems to be 
dominated rather by the longest binding length 
among the possible coalescence:
The mean value can roughly be estimated by 
the 'diagonal lengths' of four-body trapezoids, 
as shown in Fig.~\ref{tetramer-pattern} of main text. 
Taking the center of gravity of each molecule
as the vertices of trapezoids, the 'diagonal lengths' 
can be defined as the square root of the area of a trapezoid, 
which is dominated rather by the longest 
distant binding pair. 
Estimating the possibility weight for each 
trapezoid as the Boltzmann weight with 
the sum of the binding pair energies, $\Delta E$,
then the thermal averaging over the 'diagonal lengths', 
$l_{\rm FF}=3.9$\AA{} (face-to-face), 
$l_{T}=5.1$\AA{} (T-shape), and 
$l_{\rm EE}=6.0$\AA{}~\cite{2008BLU} (edge-to-edge), 
gives an improved estimate of $\bar R_{\rm tetra} = 5.0$\AA{}, 
getting closer to the experimental estimation of $R_{\rho} \sim 5.3$\AA{}. 

\par
The discrepancy still left would further be reduced by
considering the higher order clustering as well as 
the atomic vibration at finite temperature,~\cite{2016NAK} 
but the present simple idea about four-body trapezoids
seems quite successful.

\par
The above scheme also works for CHS,
as shown in the main text (See ``Validation of equilibrium properties'').}
For CHS, we can directly estimate the binding energy, $\Delta E$,
and the equilibrium binding length, $R_\mathrm{e}$,
by fitting the data using an equivalent 
form of Eq.~\eqref{lj1},
\begin{equation}
  \label{lj2}
U(R) = \Delta E \left[
2\left(\frac{R_e}{R}\right)^{6} 
- \left(\frac{R_e}{R}\right)^{12}
\right] \ ,
 \end{equation}
as summarized in Table~\ref{table:ng_table} of main text.
Note that we can also estimate $\Delta E$ and 
$R_\mathrm{eq}$ 'after' the fitting $(C_6,C_{12})$ 
first by Eq.~\eqref{lj1} in Appendix D, 
but this is not a good idea for DMC 
because the error propagation for statistical 
noises during the further transformation 
to $\Delta E$ and $R_e$ loses
the accuracy of estimates.
Fitting curves well describe the dependence
around equilibrium lengths, 
as shown in Fig.~\ref{ljfitting} of main text.
For Type-B and C with shorter-ranged exponents,
it is not rigorously validated to use the LJ potential
because its functional form assumes the $1/R^6$ asymptotic behavior.
We use it, however, under such a limited reason 
just to get possible estimates of 
$\Delta E$ and $R_\mathrm{e}$ even for Type-B and C,
as given in Table~\ref{table:two-body} of main text.

\par
We obtained $R_{\rho}=6.8$ \AA{} from
experimental values of
the molecular weight ($180.61$ g/mol) 
and density ($0.97$ g/cm$^3$ at $T = 298.15$K),
which fairly reasonably 
drops within the binding lengths of Type-A to C.
Similar to the benzene case,
the simple thermal averaging over the three configurations by the factor,
$p\sim \exp(-\Delta E/kT)$ at $T = 298.15$K
underestimates $\bar{R}_{\rm dim}\sim 4.9$ \AA{},
compared with $R_{\rho}$.
An alternative averaging
over the 'diagonal lengths' of four-body trapezoids, 
as shown in Fig.~\ref{tetramer-pattern},
gives an improved estimate, $\bar R_{\rm tetra}$,
getting closer to the experimental 
estimation, as shown in Table~\ref{table:four-body} of main text.


%

\newpage
\section{Supporting Information}

\setcounter{figure}{0}
\setcounter{equation}{0}
\renewcommand{\thetable}{S-\Roman{table}}
\renewcommand{\thefigure}{S-\arabic{figure}}
\renewcommand{\theequation}{S-\arabic{equation}}


\section{Binding curve}
For most of practical cases, we cannot expect
the molecular dimer system to be accommodated 
within the possible size to be described by 
accurate basis sets, 
such as 'triple-$\zeta$'(TZ) cc-pVTZ.
In the present case actually, 'double-$\zeta$'(DZ), 
cc-pVDZ, is the upper limit of the size 
even on the memory capacity of commercial 
super-computers.
For such a case, several schemes to correct biases 
due to less accurate basis sets are available.
Schemes for basis set superposition error 
(BSSE)~\cite{1994DUI-S} corrects
the 'unbalanced' accuracies to describe 
monomers and dimers, when they are used together 
to get binding energies.
For an implementation of BSSE, we used here 
the couterpoise method.~\cite{1970BOY-S,1996SIM-S}
Schemes of the complete basis set (CBS)~\cite{1998TRU-S}
were used to estimate an extrapolation to an enough large
basis set.

Except for CCSD(T), we applied a CBS scheme 
by Truhlar~\cite{1998TRU-S}
to get the corrected binding energy,
$\Delta E^\mathrm{}_\mathrm{CBS}$, 
by the weighting as,
\begin{equation}
\Delta E^\mathrm{}_\mathrm{CBS} 
= \frac{3^\gamma}{3^\gamma - 2^\gamma} 
\Delta E^\mathrm{}_3 - \frac{2^\gamma}
{3^\gamma - 2^\gamma} \Delta E^\mathrm{}_2 \ ,
\end{equation}
where $\Delta E_{2,3}$ denote the energies
evaluated by different basis set levels.
For more reliability, we examined two 
different pairs for the correction,
'CBS': [(2,3) = (cc-pVDZ(DZ),cc-pVTZ(TZ))], 
and 
'aCBS': [(2,3) = (aug-cc-pVDZ(aDZ),aug-cc-pVTZ(aTZ))].
The exponent, $\gamma$, is chosen as 
3.4 (2.2) for HF and B3LYP-GD3 (MP2) as proposed 
by Truhlar {\it et al.},~\cite{1998TRU-S}
which is reported to be working well for 
non-covalent systems.~\cite{2006SPO-S}

For CCSD(T), the calculation was too costly to
be done with larger basis sets other than 
cc-pVDZ level, making Truhlar's scheme 
not applicable to this case.
Instead, we hence employed Sherrill's scheme~\cite{2004SIN-S},
\begin{equation}
\label{bsse}
\Delta E^\mathrm{CCSD(T)}_\mathrm{CBS}
\approx \Delta E^\mathrm{MP2}_\mathrm{CBS} + \left(
\Delta E^\mathrm{CCSD(T)} - \Delta E^\mathrm{MP2}
\right)_\mathrm{cc-pVDZ} \ ,
\end{equation}
in which the extrapolation can be estimated
only within a basis set, but assisted by 
further MP2 evaluations.
\begin{figure*}[htb]
\begin{center}
  \includegraphics[scale=0.85]{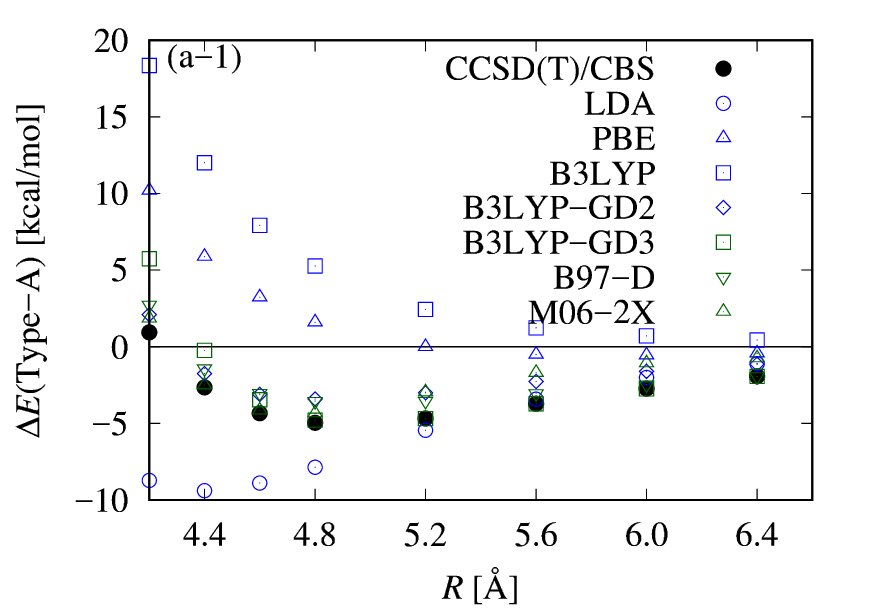}
  \includegraphics[scale=0.85]{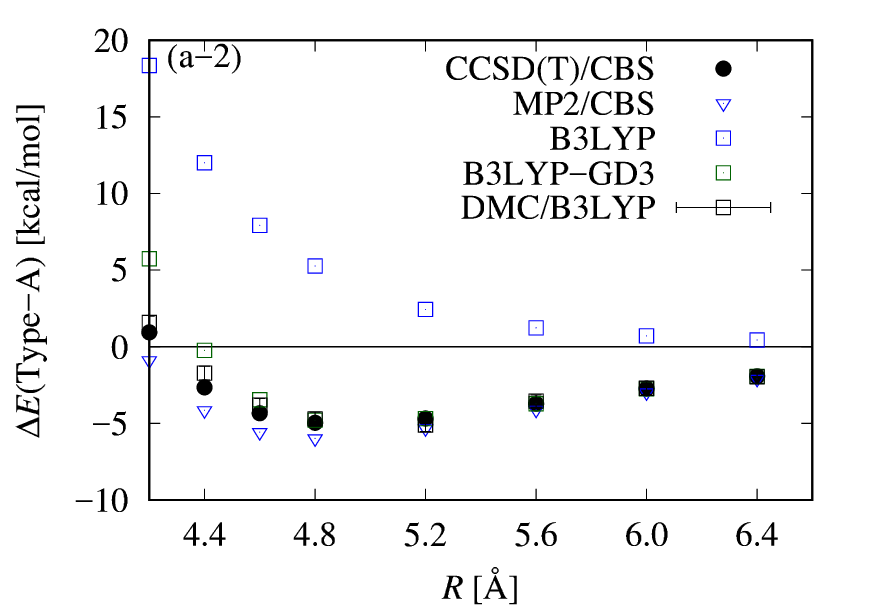}
  \includegraphics[scale=0.85]{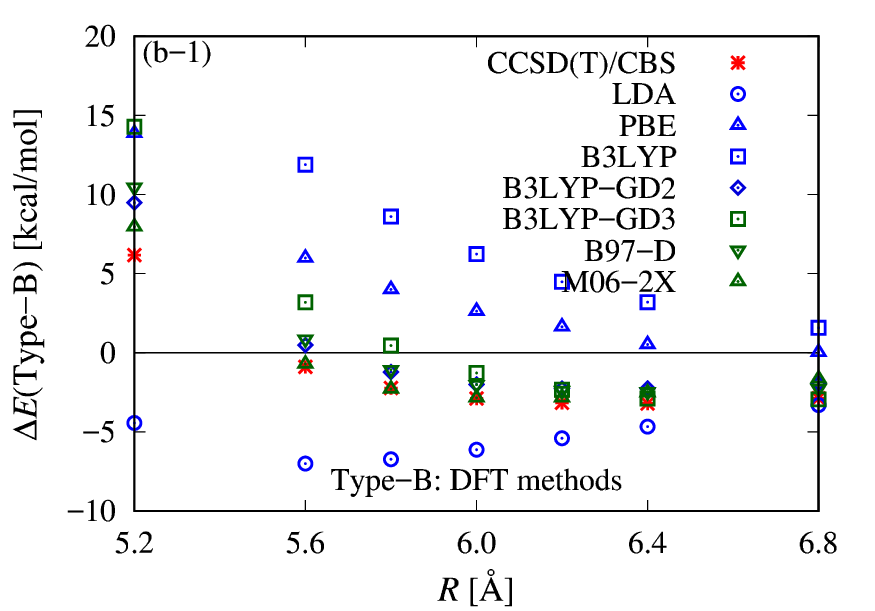}
  \includegraphics[scale=0.85]{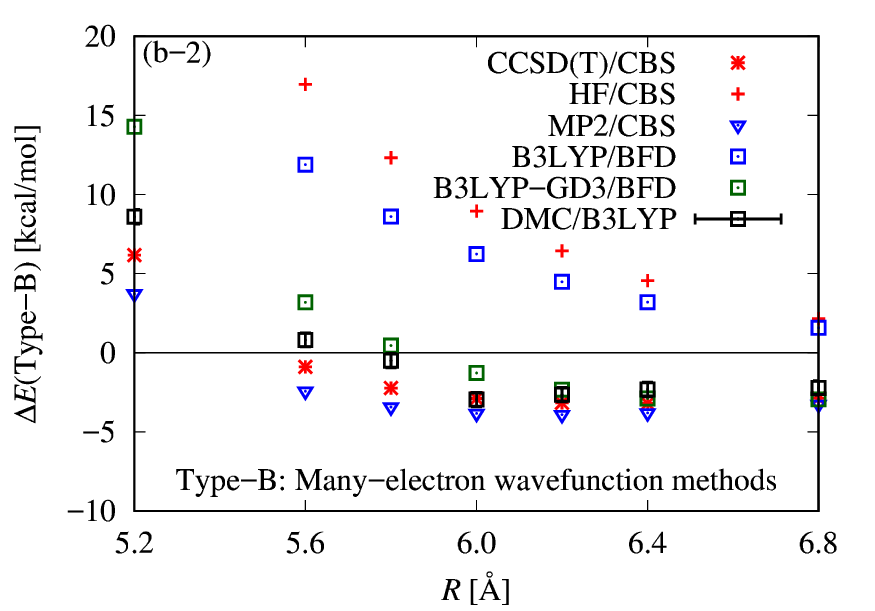}
  \includegraphics[scale=0.85]{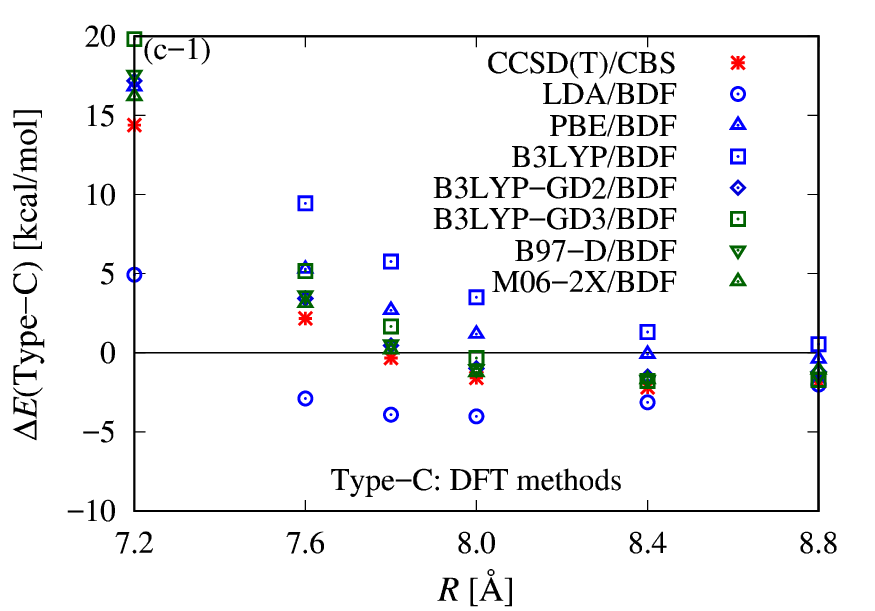}
  \includegraphics[scale=0.85]{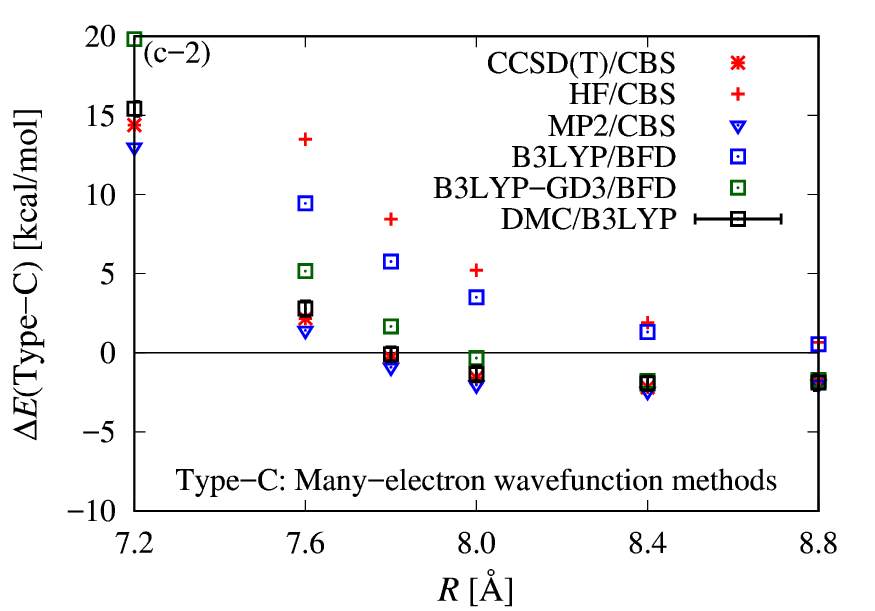}
\end{center}
\caption{
Binding curves for Sandwich (Type-A) [panels (a)], T-Shape (Type-B) [panels (b)], 
and Parallel (Type-C) [panels (c)], evaluated by several 
DFT [panel (a/b/c-1)] and correlated methods [panel (a/b/c-2)].
Results are corrected by CBS scheme if applicable.
DMC results are evaluated using B3LYP/BDF-VTZ fixed nodes.
}
\label{pesall}
\end{figure*}
\begin{figure*}[htb]
\begin{center}
\includegraphics[scale=0.85]{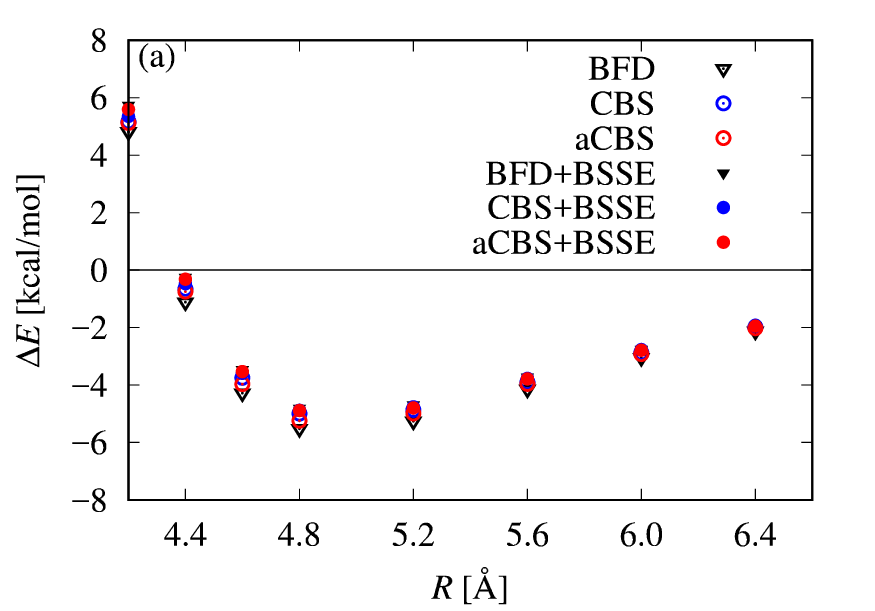}
\includegraphics[scale=0.85]{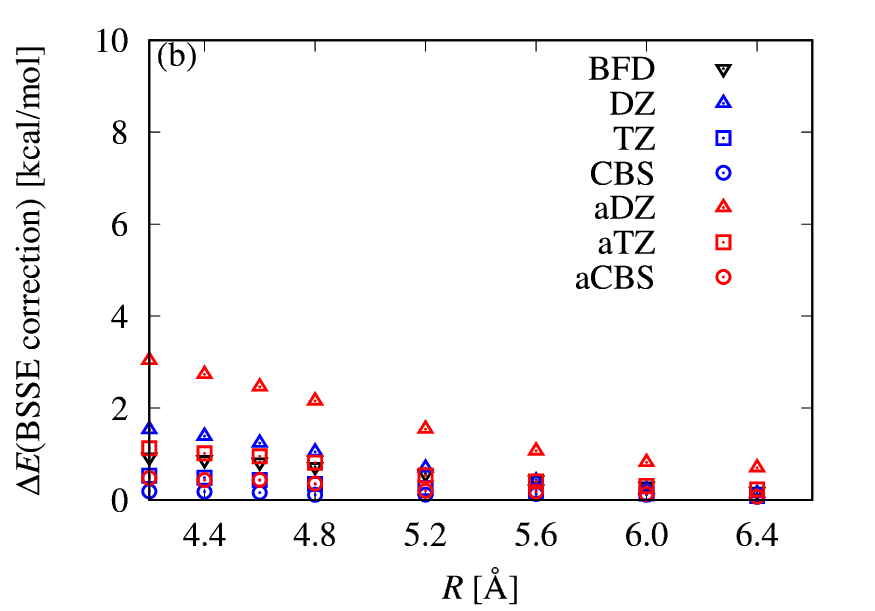}
\includegraphics[scale=0.85]{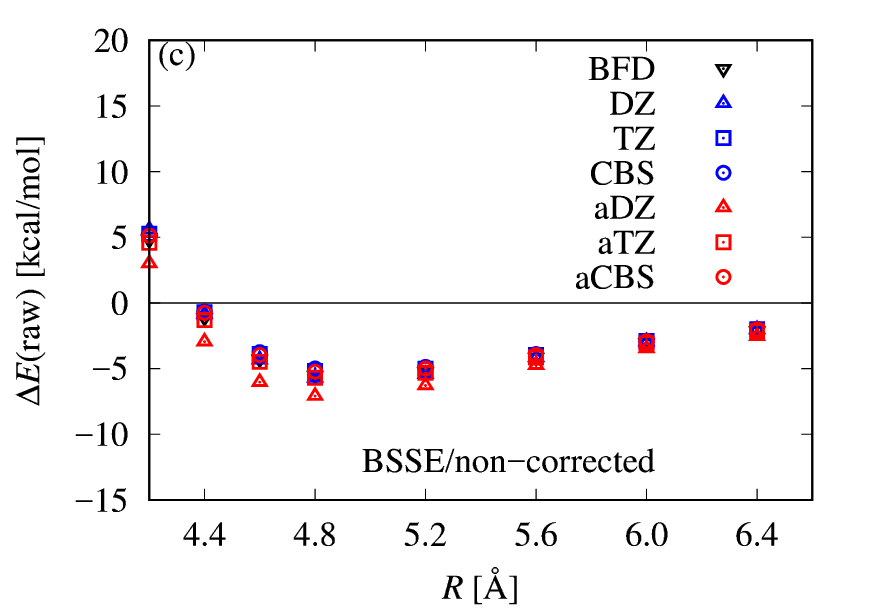}
\includegraphics[scale=0.85]{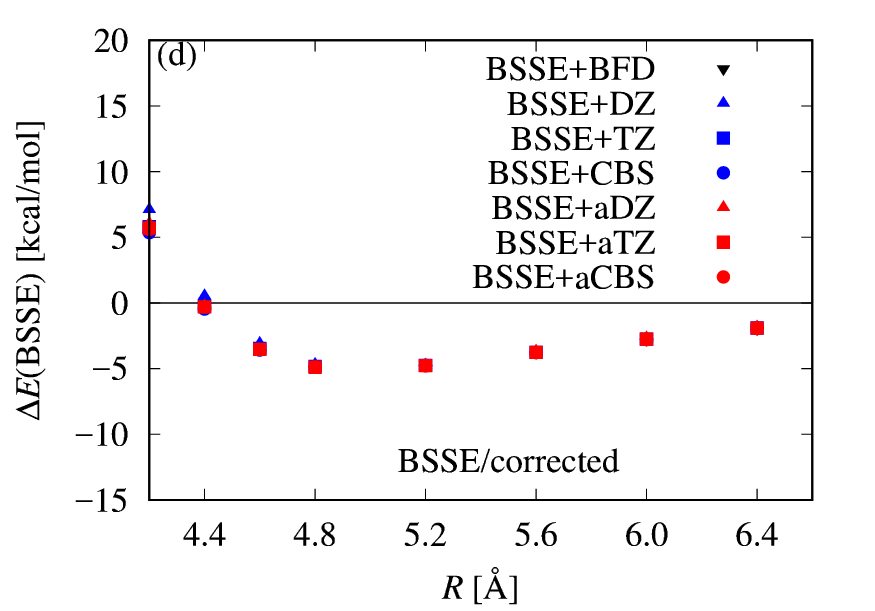}
\end{center}
\caption{
Binding curves and corrections for Type-A by B3LYP-GD3.
Panel (a) shows the final binding curves after possible
corrections, while (b)-(d) are separated contributions and
comparisons with/without BSSE corrections.
CBS (aCBS) stands for the CBS limit estimated by 
DZ/TZ (aDZ/aTZ) basis sets.
}
\label{b3lyp-gd3}
\end{figure*}
\begin{figure*}[htb]
\begin{center}
\includegraphics[scale=0.85]{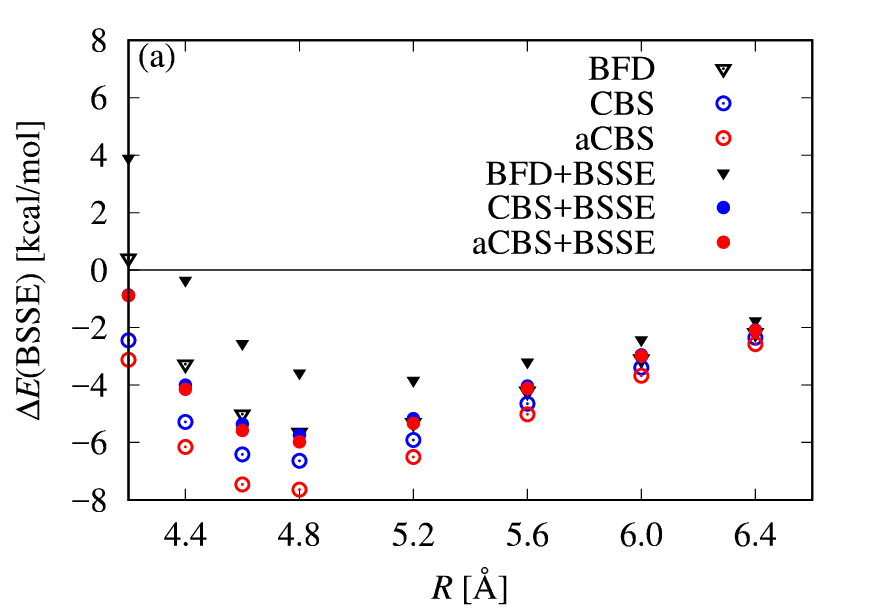}
\includegraphics[scale=0.85]{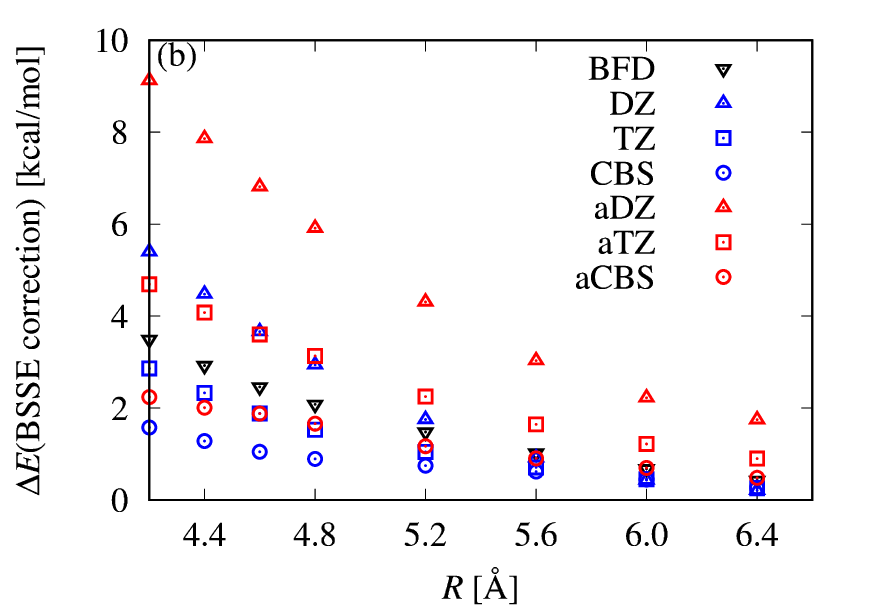}
\includegraphics[scale=0.85]{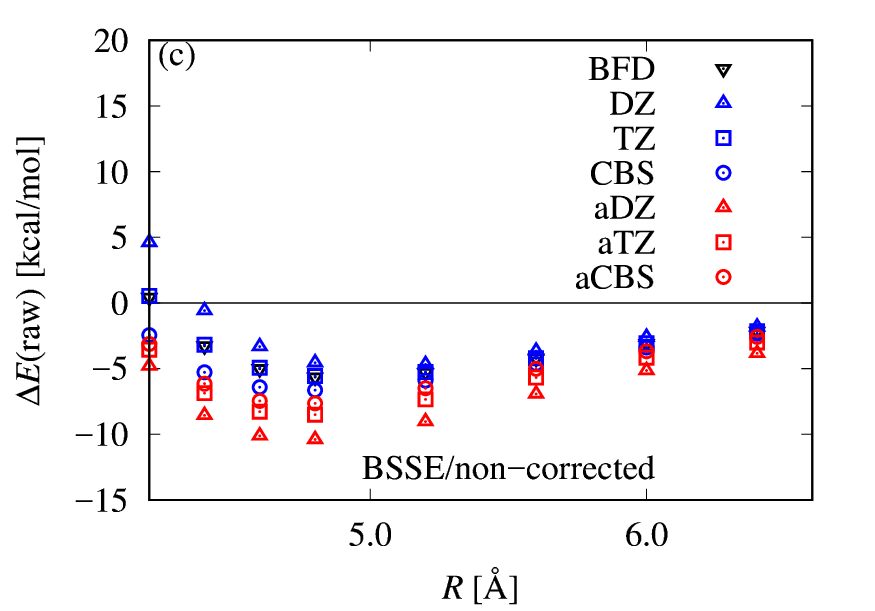}
\includegraphics[scale=0.85]{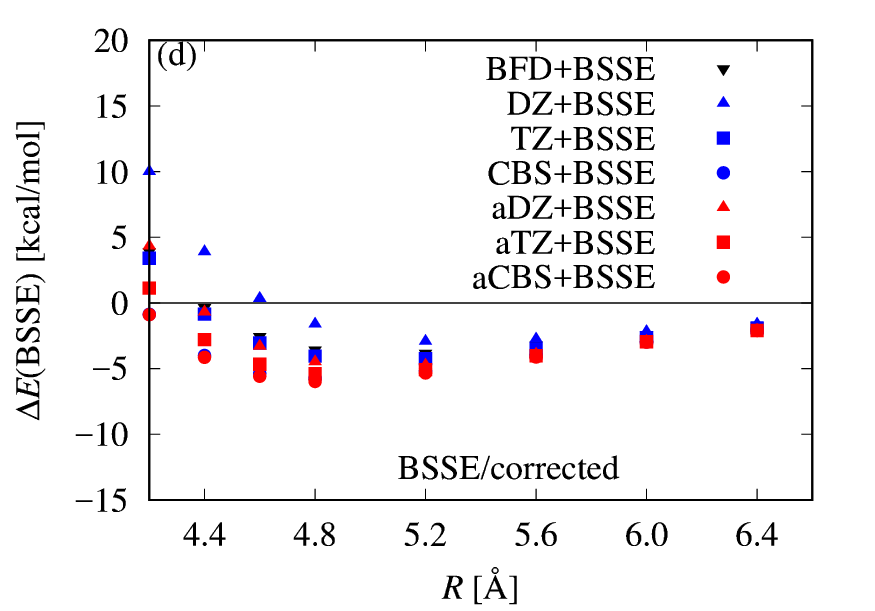}
\end{center}
\caption{
Binding curves and corrections for Type-A by MP2.
Panel (a) shows the final binding curves after possible
corrections, while (b)-(d) are separated contributions and
comparisons with/without BSSE corrections.
CBS (aCBS) stands for the CBS limit estimated by 
DZ/TZ (aDZ/aTZ) basis sets.
}
\label{mp2}
\end{figure*}
\begin{figure*}[htb]
\begin{center}
\includegraphics[scale=0.85]{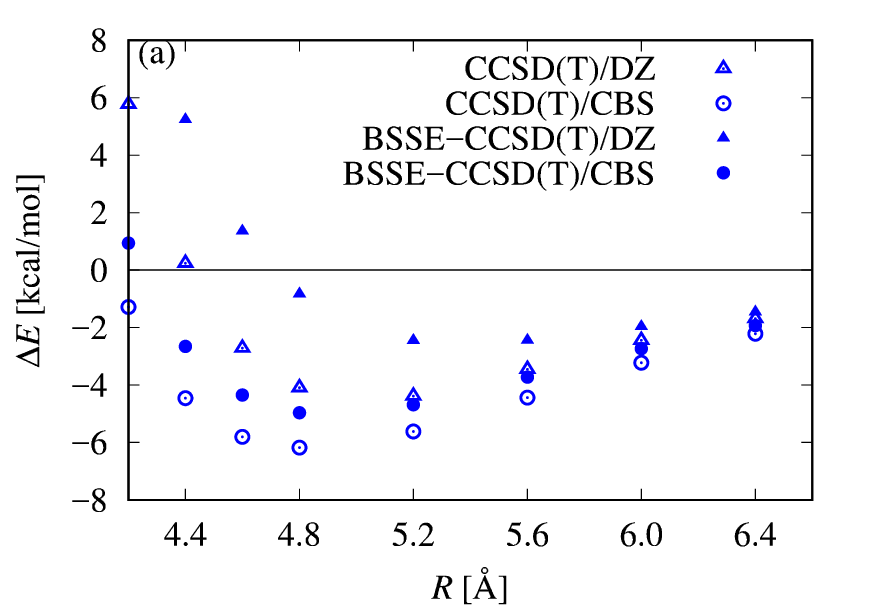}
\includegraphics[scale=0.85]{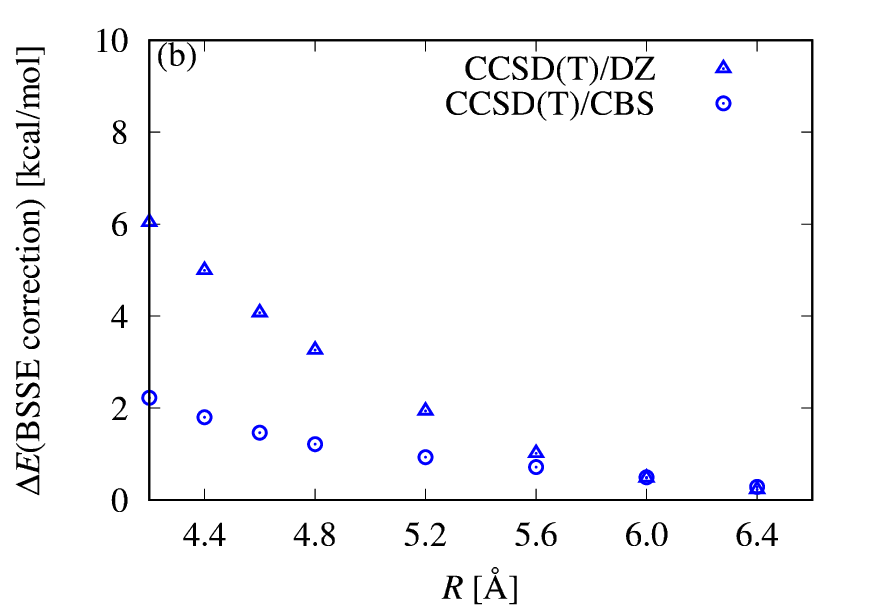}
\end{center}
\caption{
  Binding curves and corrections for Type-A by CCSD(T).
  Panel (a) shows the final binding curves after possible
corrections, while (b) shows the amount of BSSE contributions.
'CCSD(T)/DZ[CBS]' stands for the raw value without any corrections
by DZ basis sets [CBS limit], while 'BSSE-CCSD(T)/DZ[CBS]' means
that with BSSE corrections.
}
\label{ccsdt}
\end{figure*}

\vspace{2mm}
\vspace{2mm}
Binding curves of Sandwich (Type-A), T-shape (Type-B), and Parallel (Type-C)
dimer configurations are shown in Fig.~\ref{pesall}.
For SCF and correlated methods, we used Gaussian 09~\cite{2009FRI-S} and
the corresponding input files are attached to the end of this document.
For DMC/B3LYP results, all the total energies are given in the next section.
Comparisons of binding curves with/without corrections
are summarized in Figs.~\ref{b3lyp-gd3},~\ref{mp2} 
and~\ref{ccsdt} for SCF (HF and DFT in the following context), 
MP2, and CCSD(T), respectively.
In each figure, panel (a) represents the 
final results, while (b)-(d) show separated 
contributions to (a): 
In panels (b), the amount of BSSE corrections 
is found to increase as binding lengths gets shorter, 
because of the more overlapping, as expected.
We see the more accurate basis sets used, 
the smaller the amount of the BSSE correction.
Comparing panels (a) among the methods, we can see 
that correlated methods such as MP2 and CCSD(T) 
gives almost twice larger BSSE corrections
than SCF methods.
Comparison between panel (c) and (d) within each 
figure, we see that 
the dependence on basis sets gets weakened 
when BSSE corrections are applied.
For SCF (Fig.~\ref{b3lyp-gd3}), 
the dependence seems almost completely disappeared, 
while for MP2 (Fig.~\ref{mp2}) there still
remains the dependence especially on the 
predictions of the binding length. 
In SCF methods, the energy approaches to the 
CBS limit always from bottom, while in MP2 
it is alternating, namely the energy by DZ/TZ
is above the limit but below by more 
improved basis set, aDZ/aTZ.
In Fig.~\ref{b3lyp-gd3} (a), we can also confirm 
that the BFD-VTZ result is quite close to 
the CBS limit, supporting a confidence about the 
present DMC using this basis set.

\vspace{2mm}
The larger BSSE corrections for correlated methods 
(MP2 and CCSD(T)) than those for SCF. This can be explained 
as follows: 
Since the correction should be zero in the CBS limit, 
the amount of the correction would be a measure 
how far the basis set adopted is from CBS limit 
in the sense of the accuracy in each method.
Suppose a basis set being sufficient to 
describe occupied orbitals used in SCF, 
but it is not the case also for further 
unoccupied orbitals in general, which are used in 
correlated methods such as CCSD(T) or MP2.
The BSSE corrections with the same basis set 
is then getting larger for correlated methods 
than for SCF.

\section{Nodal surface dependence and time-step errors}
\begin{figure*}[htb]
\begin{center}
\includegraphics[scale=0.85]{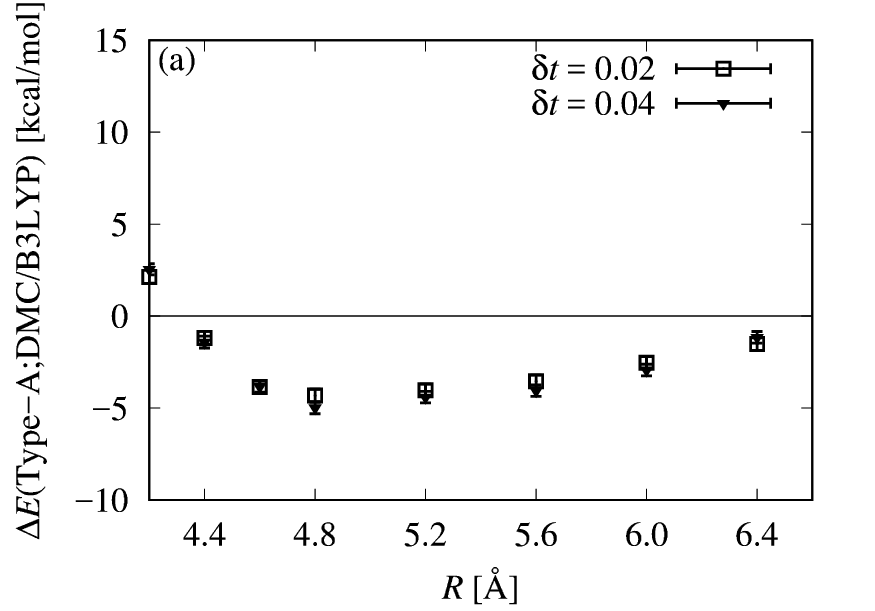}
\includegraphics[scale=0.85]{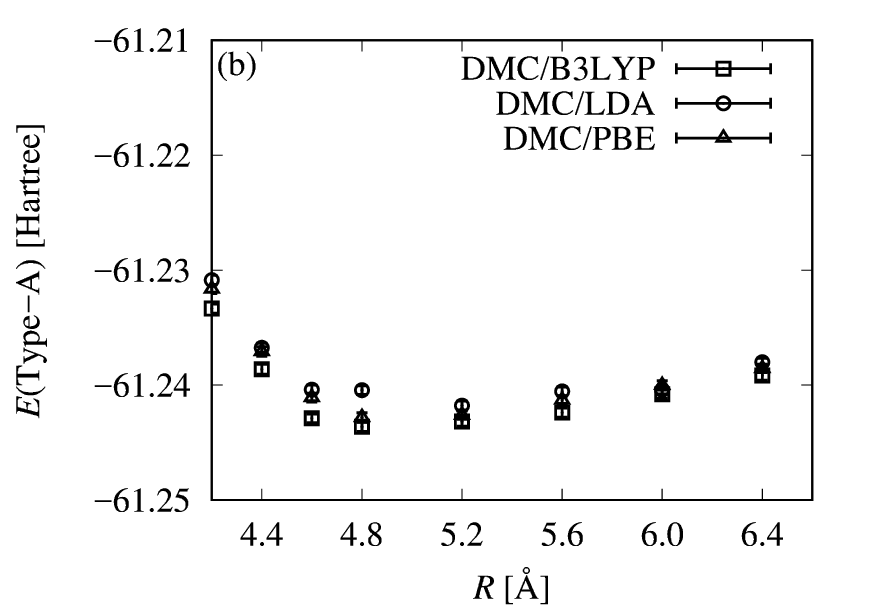}
\includegraphics[scale=0.85]{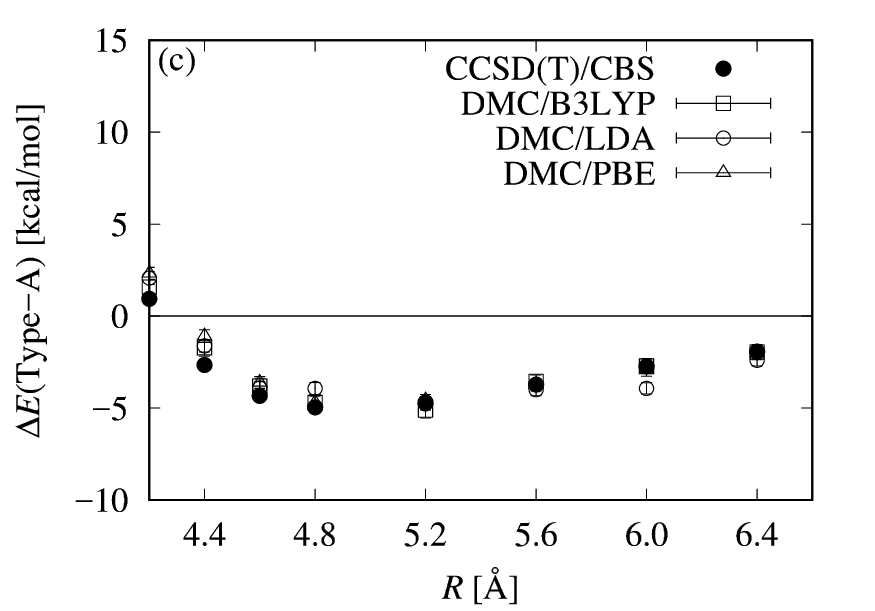}
\end{center}
\caption{
Time step dependence (left panel) and the 
nodal surface dependence (right panel) 
of the DMC binding curve for Type-A.
For the left panel, DMC/B3LYP results 
are shown, which is variationally best
in the right panel.
}
\label{dmc}
\end{figure*}
For the present DMC results, 
we have to examine the biases due to 
the approximations we applied, namely, 
the time-step approximation~\cite{1993UMR-S}
and the fixed-node approximation.~\cite{1982REY-S}
In the sense of finite discretization
of propagations, the smaller time-step, 
$\delta t$, would be
reliable, but too small step cannot 
achieve such a random walk covering 
over the sampling space within a limited number 
of steps by a tractable computation.
Fig.~\ref{dmc} shows the time-step ($\delta t$) 
dependence of the DMC binding curves, evaluated 
for Type-A using B3LYP nodal surfaces.
The curves seem to be converging within errorbars,
justifying the present choice of $\delta t = 0.02$
with enough high acceptance ratio being more than 99.5\%.

\vspace{2mm}
The dependence 
of the binding curve (Type-A/$\delta t = 0.02$)
on the nodal surfaces is shown in
Fig.~\ref{dmc} (b) in absolute energy values, 
from which we can identify which nodal surface 
gives the variationally best estimation.{\cite{1982REY-S}}
We noted that $T$-move scheme{~\cite{2006CAS-S}} is
used in the present study to preserve 
the variational principle even under the
locality approximation{~\cite{1991MIT-S}} 
for pseudo potentials.
The choice of the nodal surfaces hardly
changes the global shape such as the binding 
length.
We also see that B3LYP nodes gives the
variationally best description.
Though there is still not enough convincing 
explanations, B3LYP nodes for DMC are 
also reported as the best for several other
systems.~\cite{2013HON-S,2012HON-S,2010KOL-S,2012MAN-S}

\vspace{2mm}
All the DMC/B3LYP total energies with statistical errors
are listed in Tables~\ref{table:total}. To compute binding energies,
the reference is chosen at $R = 12.0$~\AA{} for all the cases.
They are depicted in Fig.~\ref{pesall}. In this table, under bars
in $R$ values indicate that the corresponding binding energies
are used for the log-fitting shown in Figure 8 (see Appendix D).
\begin{table*}[htb]
\begin{tabular}{lllllllll}
$R$ & Sandwich  & Error  & 
$R$ & T-Shape   & Error  & 
$R$ & Parallel  & Error  \\
\hline
4.2 &	-61.23274 &	0.00050 &	5.2 &	-61.22226 &	0.00053 &	7.2 &	-61.21068 &	0.00049 \\
4.4 &	-61.23801 &	0.00059 &	5.6 &	-61.23467 &	0.00046 &	7.6 &	-61.23080 &	0.00053 \\
4.6 &	-61.24133 &	0.00053 &	5.8 &	-61.23676 &	0.00050 &	7.8 &	-61.23540 &	0.00049 \\
4.8 &	-61.24275 &	0.00053 &	6.0 &	-61.24068 &	0.00051 &	8.0 &	-61.23738 &	0.00056 \\
5.2 &	-61.24341 &	0.00052 &	6.2 &	-61.24018 &	0.00050 &	8.4 &	-61.23834 &	0.00048 \\
5.4 &	-61.24324 &	0.00052 &	6.4 &	-61.23965 &	0.00052 &	\underbar{8.8} &	-61.23824 &	0.00060 \\
5.6 &	-61.24091 &	0.00049 &	\underbar{6.8} &	-61.23950 &	0.00048 &	\underbar{10.0} &	-61.23660 &	0.00051 \\
6.0 &	-61.23959 &	0.00048 &	\underbar{7.0} &	-61.23868 &	0.00052 &	\underbar{11.0} &	-61.23578 &	0.00047 \\
\underbar{6.4} &	-61.23838 &	0.00049 &	\underbar{8.0} &	-61.23775 &	0.00050 &	12.0 &	-61.23524 &	0.00056 \\
\underbar{6.6} &	-61.23798 &	0.00051 &	12.0 &	-61.23596 &	0.00052 &	     &		    &           \\
\underbar{7.0} &	-61.23718 &	0.00046 &	     &		  &		&	     &              &           \\
12.0 &	-61.23526 &	0.00043 &            &		  &		&	     &              &           \\
 \hline
\end{tabular}
\caption{
Total energies with statistical errors for Sandwich (Type-A), T-Shape (Type-B),
and Parallel (Type-C) dimer configurations, evaluated from DMC/B3LYP with
$\delta \tau = 0.02$. All energies and lengths are given in hartrees and angstroms, respectively.
}
\label{table:total}
\end{table*}

\providecommand{\latin}[1]{#1}
\providecommand*\mcitethebibliography{\thebibliography}
\csname @ifundefined\endcsname{endmcitethebibliography}
  {\let\endmcitethebibliography\endthebibliography}{}

\begin{figure*}[htb]
\begin{center}
\includegraphics[scale=0.85]{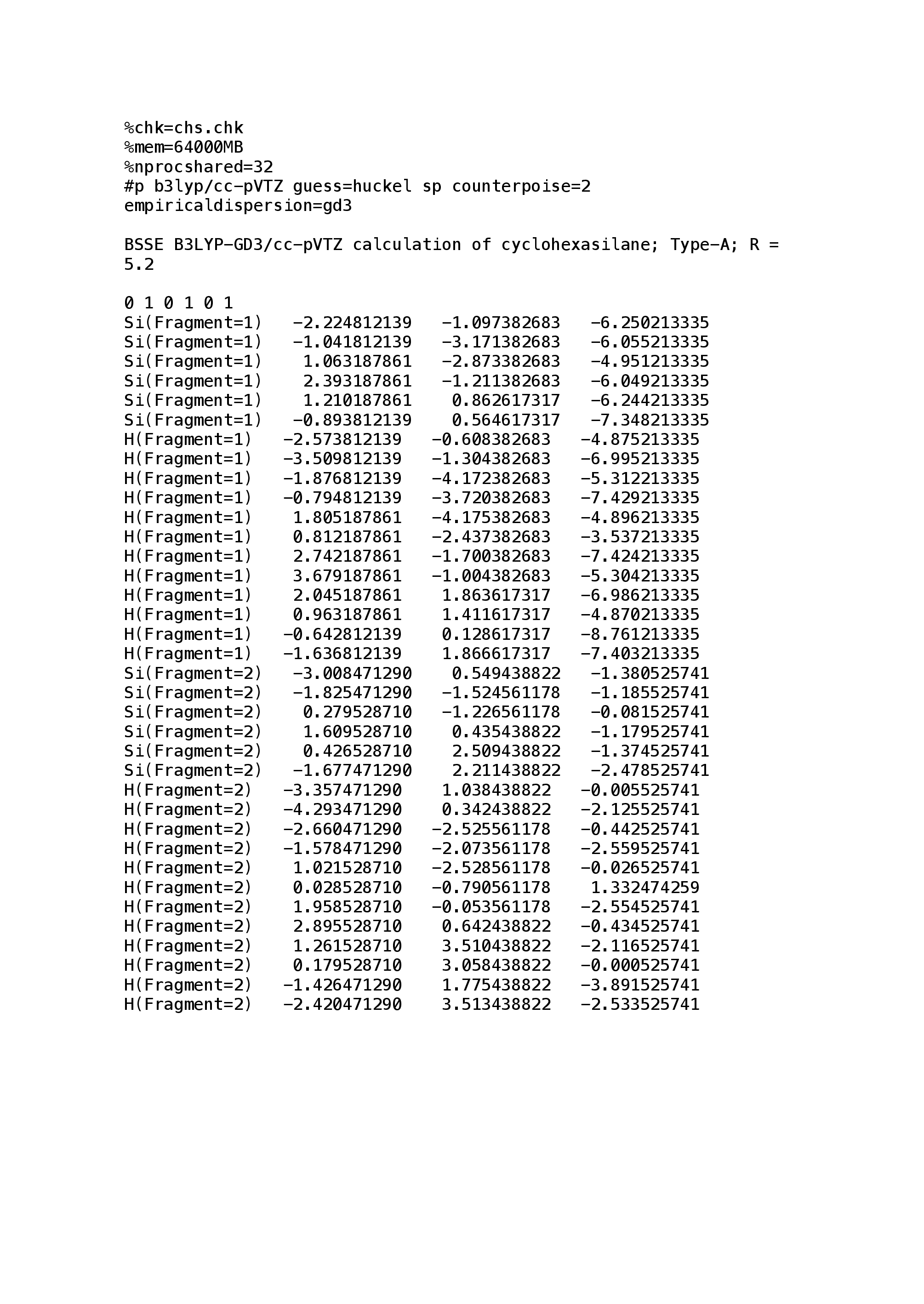}
\end{center}
\label{G09-1}
\end{figure*}
\begin{figure*}[htb]
\begin{center}
\includegraphics[scale=0.85]{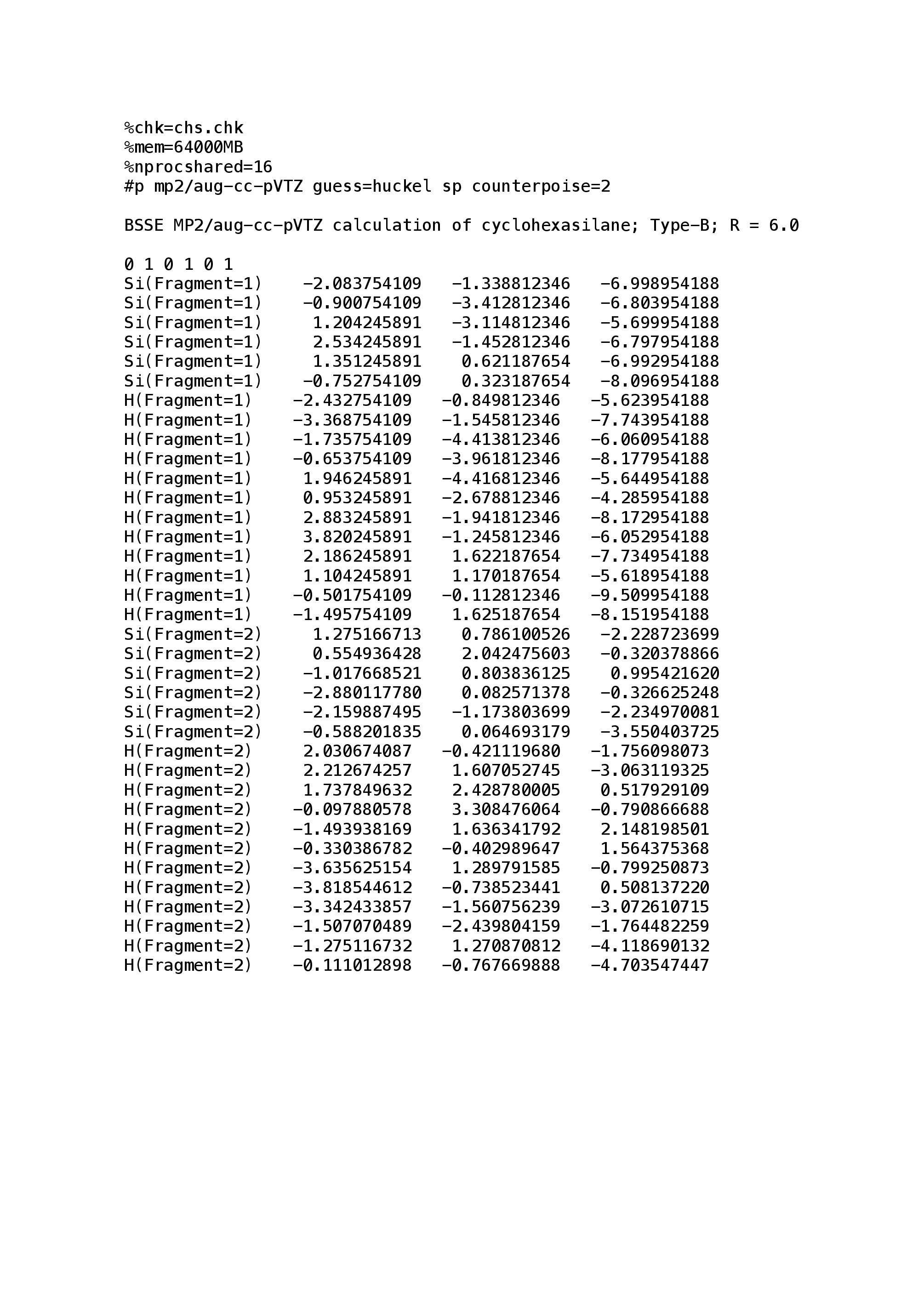}
\end{center}
\label{G09-2}
\end{figure*}
\begin{figure*}[htb]
\begin{center}
\includegraphics[scale=0.85]{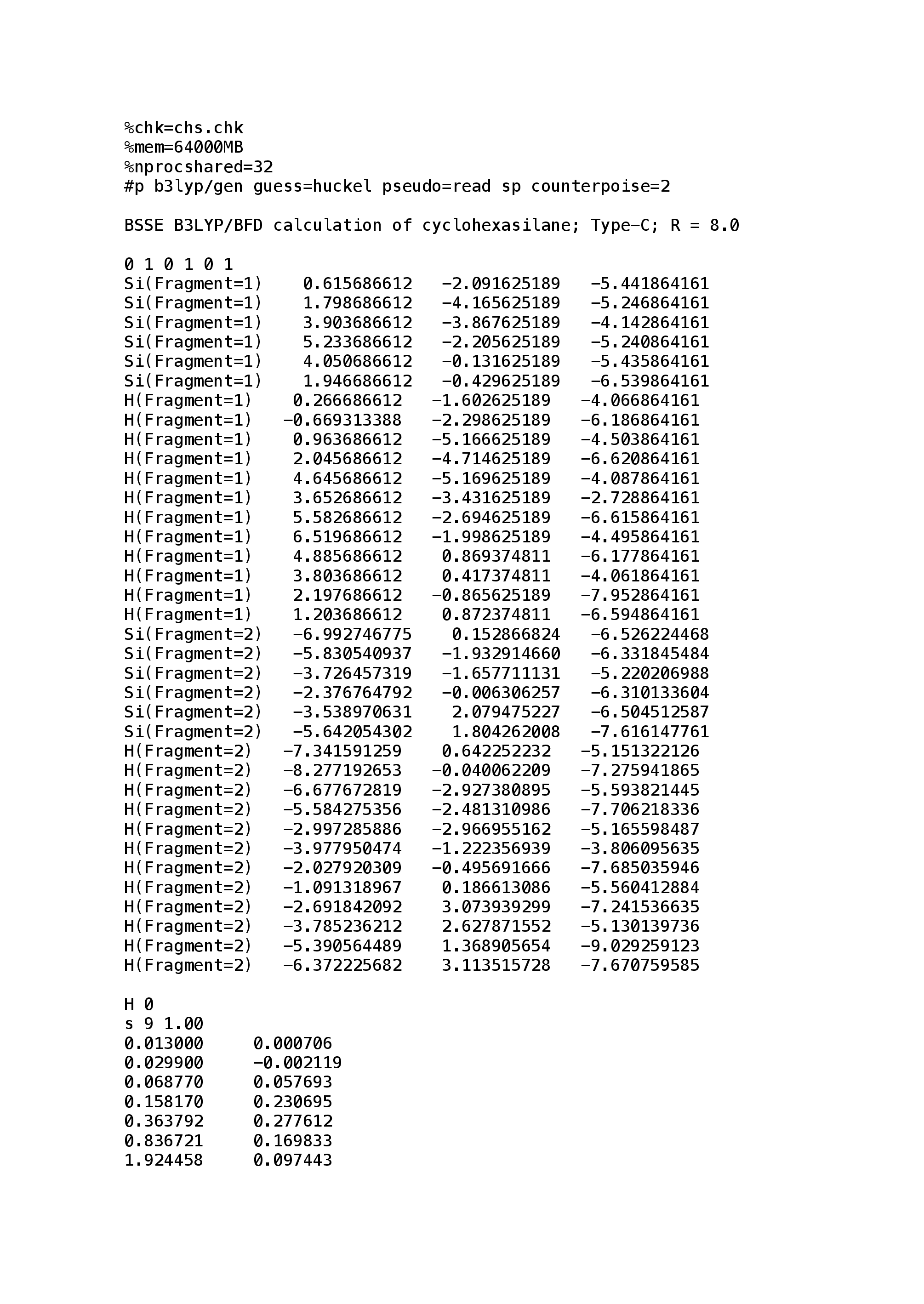}
\end{center}
\label{G09-3}
\end{figure*}
\begin{figure*}[htb]
\begin{center}
\includegraphics[scale=0.85]{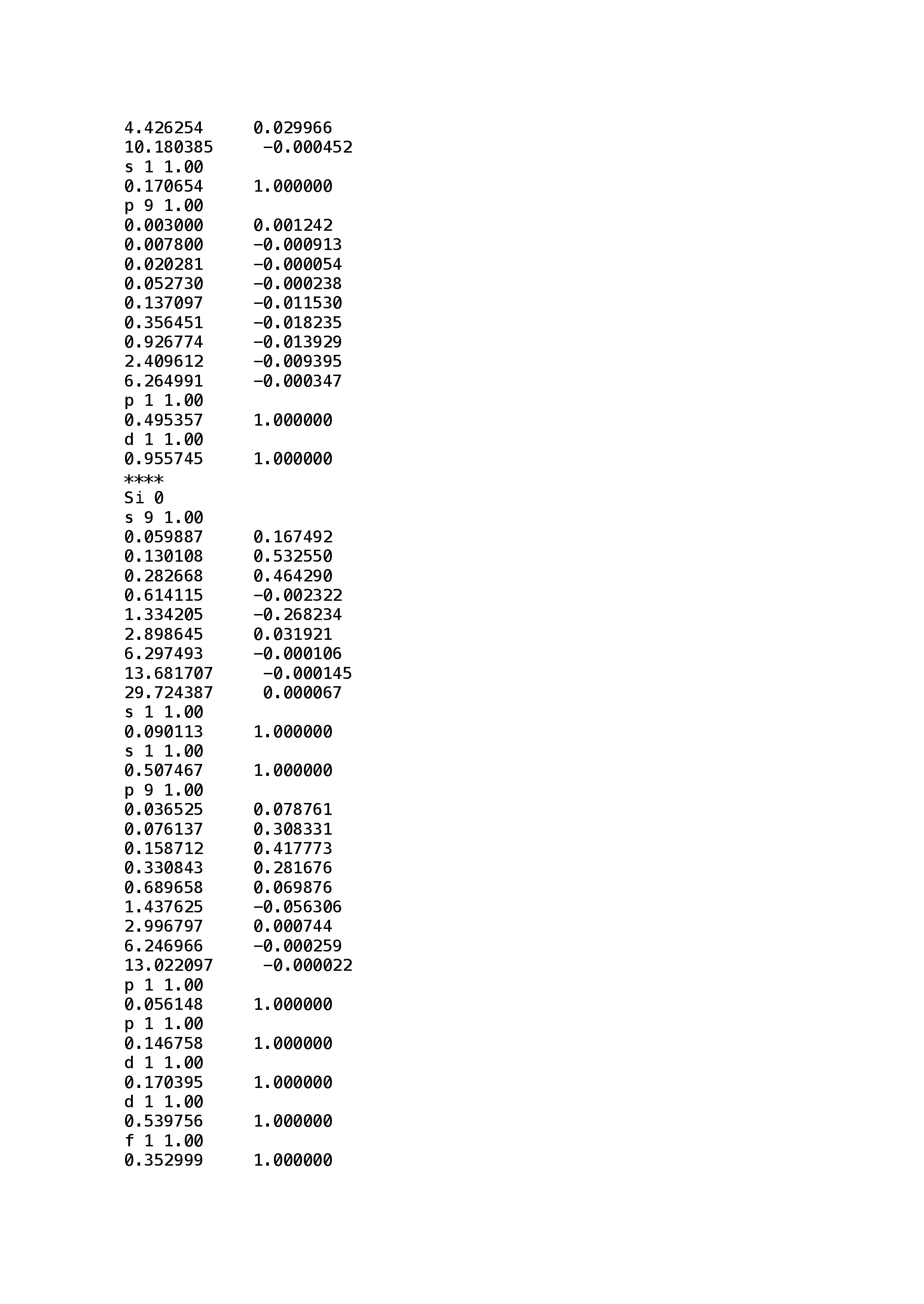}
\end{center}
\label{G09-4}
\end{figure*}
\begin{figure*}[htb]
\begin{center}
\includegraphics[scale=0.85]{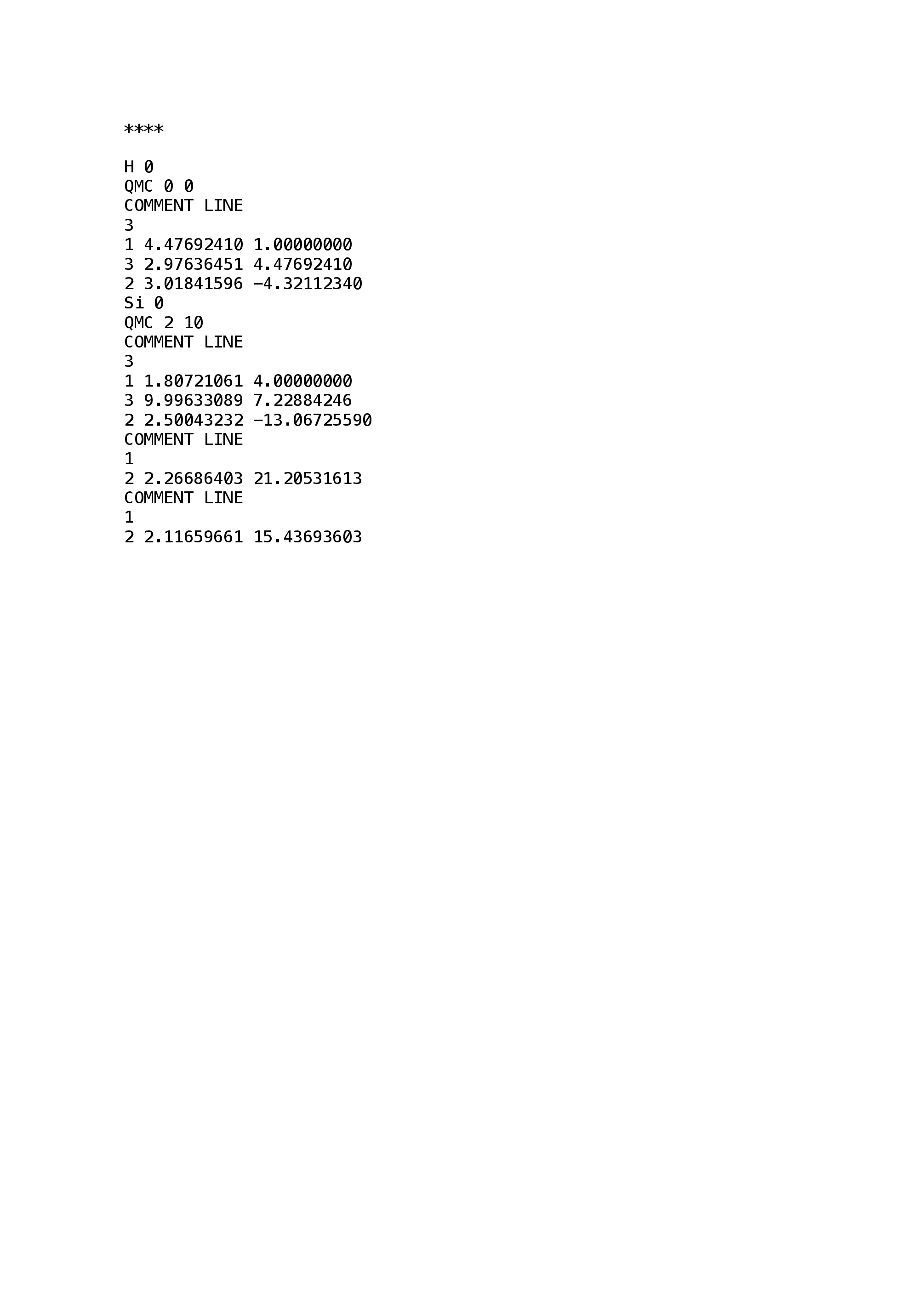}
\end{center}
\label{G09-5}
\end{figure*}

\end{document}